\documentclass[genTeX]{nrc1}   
\usepackage[pctex32]{graphicx}

\volyear{99}{2005}
\received{2004}
\accepted{2005}
\begin{document}

\title{Fundamental Physical Constants:\\ Looking from Different Angles}
\author{Savely G. Karshenboim}
\address{D. I. Mendeleev Institute for Metrology, St. Petersburg, 189620, Russia and
Max-Planck-Institut f\"ur Quantenoptik, Garching, 85748, Germany; e-mail: sek@mpq.mpg.de}

\shortauthor{Savely Karshenboim}

\maketitle

\begin{abstract}
We consider fundamental physical constants which are among a few
of the most important pieces of information we have learned about
Nature after its intensive centuries-long studies. We discuss
their multifunctional role in modern physics including problems
related to the art of measurement, natural and practical units,
origin of the constants, their possible calculability and
variability etc.
\end{abstract}
\begin{resume}
Nous ... French version of abstract (supplied by CJP)
   \traduit
\end{resume}

\pagebreak

\tableofcontents

\pagebreak

\section{Introduction\label{s:in}}

\begin{flushright}
\begin{minipage}{9cm}
`You needn't say ``please'' to {\em me\/} about 'em,' the Sheep
said, ...`I didn't put 'em there, and I'm not going to take 'em
away.' \centerline{\em L.C.}
\end{minipage}
\end{flushright}

There is a number of ways to understand Nature. One can approach
it with logics, with guesses, with imagination. A way
scientists and, especially, physicists address the problem is
based on a comparison of our ideas and the reality via
measurements. Experiment inspires theory and verifies theory. The
soul of the physical approach is neither logics, nor even a
quantitative approach, but a common sense. The latter is based on
centuries-long experience in investigation of Nature. It tells us
to be sceptical. It tells us that even complicated phenomena are
often based on simple pictures, and most of them allow to estimate
effects in terms of certain fundamental quantities. It tells us
that those simple pictures are good to start with but should
involve more and more details once we desire a more accurate
agreement between any of our theories and the measured reality.

Fundamental constants play a crucial role in physics in a few
different ways and we consider their significance in this paper.
However, to start the subject we need to agree on what the
fundamental constants are. And we discover a great variety of
approaches to the problem based on a particular role of a
particular constant in a specific field of physics.

Two polar points of view are related to `practical' and
`fundamental' physics.
\begin{itemize}
\item The practical view addresses the art of measurement
which makes physics to be physics. There is a number of beautiful
laws such as the Maxwell equations or the Dirac equation which
pretend to describe Nature. However, as a quantitative method of
exploring the world, physics needs some quantitative values to be
measured. That requires certain parameters enter basic
equations. We also need certain quantities to be used as units to
make proper comparisons of different results. Some of these
parameters enter a number of equations from different branches of
physics and are universal to some extent. That is a `practical'
way to define what the fundamental constants are. A very important
property of such constants is that they should be measurable. The
fundamental constants understood in such a way are a kind of
an interface to access Nature quantitatively and apply basic laws
to its quantitative description.
\item However, not every such a constant is truly fundamental. If
we, e.g., need a unit, we can consider a property of such a
non-fundamental object as the caesium atom. The approach of
fundamental physics is based on the idea that we can explain the
world with a few very basic laws and a few very basic constants. The
rest of the constants should be calculable or expressed in terms
of other constants. Such constants are our interface to really
fundamental physics but most of them have a very reduced value in
real measurements, because they are often not measurable. To
deduce their values from experiment, one has to apply
sophisticated theories and, sometimes, certain models.
\end{itemize}

\begin{figure}[hbtp]
\begin{center}
\includegraphics[width=0.7\textwidth]{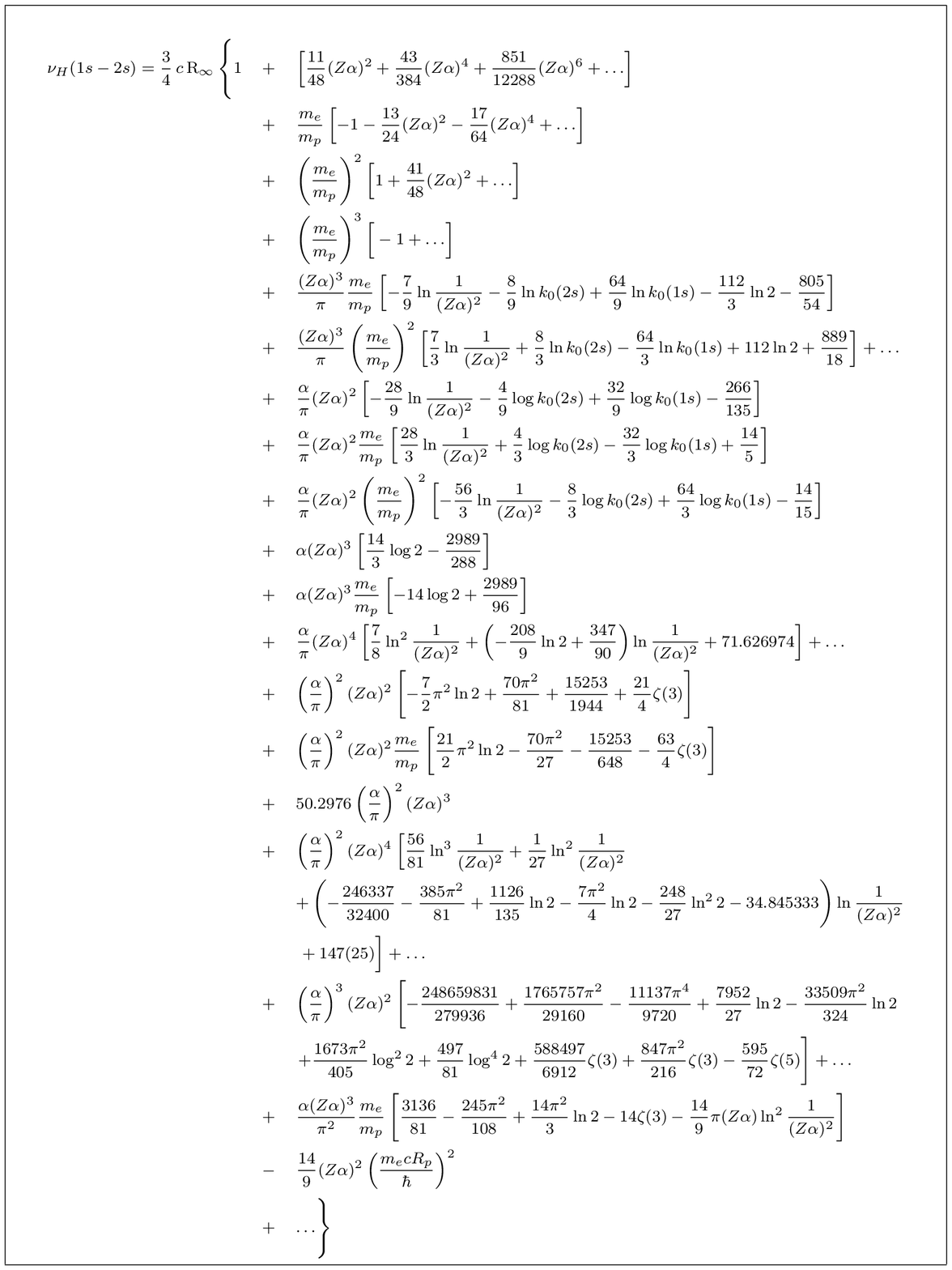}
\end{center}
\caption{A relation between the $1s-2s$ transition frequency
$\nu_H(1s-2s)$ and the Rydberg constant $R_\infty$. A correction
for a difference between the center of gravity of the $1s$ and $2s$
hyperfine multiplets and their triplet component is not included.
\label{F:nury}}
\end{figure}

A good illustration of a difference between these two approaches
is the situation with the Rydberg constant
\begin{eqnarray}\label{rydef}
R_\infty &=& \frac{\alpha^2m_ec}{2h}\nonumber\\
&=& \frac{e^4m_e}{8\epsilon_0^2h^3c}\;,
\end{eqnarray}
which is expressed in a simple way in terms of certainly more
fundamental quantities. However, this exactness is rather an
illusion, because the constant is not measurable in a direct way.
The most accurately measured transition in hydrogen is the triplet
$1s-2s$ transition (see, e.g., \cite{Niering})
\begin{equation}\label{1s2s}
\nu_H(1s-2s,\; F\!=\!1)
=2\,466\,061\,102\,474\,851(34)\;\mbox{Hz}~~~~[1.4\times10^{-14}]
\end{equation}
and it may be only approximately related to the Rydberg constant
(see Fig.~\ref{F:nury}). To obtain its value, one has to apply
quantum electrodynamics (QED) theory and perform some additional
measurements. At the present time, the
fractional accuracy (a number in squared brackets) in the
determination of this constant \cite{codata}
\begin{equation}\label{ry10}
R_\infty =10\,973\,731.568\,525(73)\; \mbox{\rm
m}^{-1}~~~~[6.6\times10^{-12}]
\end{equation}
is much lower than that of the measurement of $\nu_H(1s-2s)$ (cf.
(\ref{1s2s})). One might indeed redefine the Rydberg constant in
some other way, e.g.
\begin{equation}
\widetilde{R} = \frac{4}{3\,c}\,\nu_H(1s-2s)\;,
\end{equation}
making the accuracy of its determination higher. However, a
relation with more fundamental constants would be more complicated
and not exactly known. The practical and fundamental approaches
cannot easily meet each other because we can very seldom both
exactly calculate and directly measure some quantity, which has a
certain non-trivial meaning. A choice between the practical and
fundamental options is a kind of trade-off between measurability and
applicability, on one side, and calculability and theoretical
transparency, on the other.

There is a number of approaches which lie between the two
mentioned above. For example, quantum electrodynamics (QED) at the
very beginning of its development met a problem of divergencies in
simple calculations. A response to the problem was the idea of
renormalization, which states that QED theory should express
observable values (such as, e.g., the Lamb shift in the hydrogen
atom or the anomalous magnetic moment of an electron) in terms of
certain observable properties (such as the charge and mass of an
electron). The measurable charge and mass are definitely more
fundamental than most of `practical' constants such as caesium
hyperfine constant. However, they are certainly less fundamental
than similar quantities defined at the Planck scale. Such an
approach is in formal sense not an {\em ab initio\/} calculation
of a quantity under question (e.g., the Lamb shift), but rather an
{\em ab initio\/} constraint on observable quantities (the energy
shifts, the charges and the masses).

The point which definitely unifies all approaches is that the
fundamental constants are such dimensional or dimensionless
quantities which are fundamentally important to understand,
investigate and describe our world. However, the {\em importance\/}
is often understood differently. In a finite-size paper it is not
easy to consider the whole range of problems related to the fundamental
constants and part of discussion missing here can be found in another
recent review of the author \cite{UFN} (see also \cite{ACFCi}).

\section{Physical Constants, Units and Art of Measurement\label{s:art}}

\begin{flushright}
\begin{minipage}{9cm}
Speak in French when you can't think of the English for a thing.
\centerline{\em L.C.}
\end{minipage}
\end{flushright}

When one does a measurement, certain units should be applied to
arrive at a quantitative result. A measurement is always comparison
and in a sense we deal with dimensionless quantities only.
`Mathematically', this point of view is true, however, it is
counterproductive. To compare two similar quantities measured
separately, we have to go through a number of comparisons. Instead
of that, it has been arranged to separate a certain part of
comparisons and use them to introduce {\em units\/}, certain
specific quantities applied worldwide for a comparison with
similar quantities under question. The units (or a system of
units) find its endower as a coherent system of certain
universally understood and legally adopted measures and weights
which can be used to measure any physical quantity.

One should not underestimate problems of measurements. Access to
quantitative properties of Nature is a crucial part of physics and
it is a problem of fundamental importance to improve and extend
our accessibility.

Since the very discovery of the world of measurable quantities, we
used natural units. But their degree of naturality was different.
We started with values related to our essential life:
\begin{itemize}
\item parameters of human beings;
\item parameters of water, the most universal substance around us;
\item parameters of Earth itself;
\item parameters of Earth as a part of the Solar system.
\end{itemize}
This approach had been followed until the introduction of the {\em
metric system\/} two centuries ago, and, in fact, the very metric
system was originally based on properties of Earth: the
metre\footnote{There are two different spellings for this term:
the {\em meter\/} is used in USA, while the {\em metre\/} is used
in UK and most of other English speaking countries and in
international literature (see, e.g., \cite{SI}). The latter is also
traditionally used in metrological literature.} was defined in
such a way that a length of a quadrant (a quarter of meridian) of
a certain meridian was equal to $10\,000\;$km (exactly); the
second was obviously defined by a day and a year; the
gram was then understood as a mass of one cubic centimetre of
water and so on.

The metric treaty was signed in 1875 in Paris. Since then we have
changed contents of our units but tried to keep their size. Few
changes took place after the SI system was adopted in 1960 (`SI'
means {\em Syst\`eme International d'Unit\'{e}s\/} ---
International System of Units). The latest version is presented in
an CIPM\footnote{CIPM is the International Committee for Weights
and Measures.} brochure \cite{SI}. For example, the SI unit of
length, the {\em metre\/}, was originally defined via the size of
Earth, later was related to an artificial ruler, to a hot optical
emission line and now to the hyperfine structure interval in cold
caesium atoms
\begin{equation}
\nu_{\rm HFS}({}^{133}{\rm Cs}) = 9\,192\,631\,770 \; \mbox{Hz}~~~~{\rm (exactly)}
\end{equation}
and a fixed value
of the speed of light
\begin{equation}
c = 299\,792\,458 \;\mbox{m/s}~~~~{\rm (exactly)}\;.
\end{equation}
One also has to remember that this unit was introduced as a
substitute for numerous units based on details of the shape of a
human body, such as the {\em foot\/} and the {\em yard\/} and takes their
magnitude (in a general scale) from them
\begin{equation}
1 \; \mbox{m}  \simeq 1.1 \;\mbox{yd} \simeq  3.3 \; \mbox{ft} \;.
\end{equation}

This evolution in the definition of the metre clearly demonstrates two
great controversies of the SI system: {\em changing stability\/}
and {\em advanced simplicity\/}. First, there has been no single
SI system at all. We have seen with time since the very
appearance of the metric convention, a number of various, but
similar, systems of units. While the hierarchy and basic relations
between the units were roughly the same all over the time, the
units themselves and related standards changed drastically.
However, for practical reasons, the size of the units during
those revolutionary redefinitions was kept the same as close as
possible. Secondly, the system of units is indeed a product
created first of all by non-physicists for non-physicists. I
dare say, we, physicists, even do not care what the SI
units actually are. For example, most of us have learned about
the {\em mole\/} in a
time, when we could not recognize that excitations, binding
effects in the solid phase, kinetic energy etc would change a mass
of the sample (but indeed not a number of particles). Later, after
we have learned about all these effects, we assumed that the SI
definition is properly adjusted to them. But it is unlikely that most
of us checked the SI definitions for that. Actually, all SI
definitions come historically from non-relativistic classical
physics and similar to their appearance we have also learned them
for the first time as classical non-relativistic stuff. We do not
care about actual SI definitions partly because we do not
consider seriously the legal side of SI and due to that we believe
that we may ourselves interpret and correct SI definitions if
necessary.

Physicists serve as experts only while decisions are made by
authorities. The SI system has been created for a legal use and
trade rather than for scientific applications. Due to that, crucial
features of the SI convention should be expressed as simply as
possible. Meanwhile, these `simply defined' units should allow to
apply the most advanced physical technologies. That makes the SI
system to be a kind of an iceberg with a stable and simple visible
part, while the underwater part is sophisticated, advanced and
changes its basic properties from period to period. The changes in
the definition of the SI metre have demonstrated a general trend
in physical metrology: to use more stable and more fundamental
quantities and closely follow progress in physics. Eventually, we
want units to be related to quantized properties of natural
phenomena and most of all, if possible, to values of fundamental
constants. We already have natural definitions of the metre and
the second, and are approaching a natural definition of basic
electric units and, maybe, the kilogram.

Note, however, that a choice of units is not restricted to the
International system SI. There is a number of options. Certain
units, such as {\em universal atomic mass unit\/}, are accepted to
be used together with the SI units \cite{SI}. There is a number of
units such as the {\em Bohr magneton\/} $\mu_B=e\hbar/2m_e$ and
the {\em nuclear magneton\/} $\mu_N=e\hbar/2m_p$, which do not
need any approval since they are well-defined simple combinations
of basic fundamental constants. A number of quantities are
measured in terms of fundamental constants. For example, the
electric charge of nuclei and particles is customarily expressed
in that of the positron. Sometimes, instead of introducing units,
new values with special normalization are introduced, such as
angular momentum in quantum physics, which is equal to the actual
angular momentum divided by the reduced Planck constant $\hbar$.

Fundamental constants (as units) play a very important role
in precision measurements and in special cases. The latter
corresponds to a situation when conventional methods cannot be
applied. For example, sometimes to determine a temperature we can
not use a
thermometer properly calibrated using primary thermodynamical
standards. In some cases, when, e.g., the temperature is too high or too
low, or if we have to deal with a remote object, we may rely on
the Boltzmann distribution and measure frequency and the spectral
intensity of emitted photons. To interpret the frequency in terms
of temperature, we have to use the values of the Planck constant $h$
and the Boltzmann constant $k$.

\section{Physical Constants and Precision Measurements\label{s:prec}}

\begin{flushright}
\begin{minipage}{8cm}
`You needn't say ``exactly,'' ' the Queen remarked: `I can believe
it without that. Now I'll give {\em you\/} something to
believe.'\\
\centerline{\em L.C.}
\end{minipage}
\end{flushright}

A precision measurement is another case very closely related to
fundamental constants. As we mentioned, those constants are universal
and some may appear in measurements in different branches of
physics. That offers us a unique opportunity to verify our
understanding of Nature in a very general sense. We know that any
particular theory is an approximation. Our basic approach always
involves certain laws and certain ideas on what the uncertainty of
our consideration is. The most crucial test of the whole approach is to
check if values from different areas of physics agree with each
other.

\begin{figure}[hbtp]
\begin{center}
\includegraphics[width=0.7\textwidth]{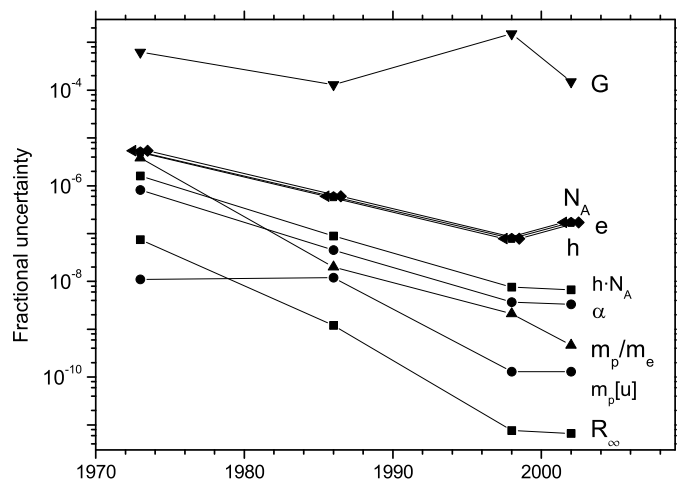}
\end{center}
\caption{Progress in the determination of fundamental constants:
the time dependence of the fractional uncertainty (see the recent paper
\protect\cite{codata} and also \cite{codataold} on earlier results
by the CODATA task group). Here, $G$ stands for the Newtonian
gravity constant, $N_A$ for the Avogadro constant and $m_p[u]$ for
the proton mass in the universal atomic mass units. \label{f:pro}}
\end{figure}

This test has been regularly performed by the {\em
CODATA\footnote{CODATA is the Committee for Data for Science and
Technology of the International Council for Science.} task group
on fundamental constants\/}, which publishes its {\em Recommended
Values of the Fundamental Constants\/} \cite{codata} (see also
previous CODATA papers \cite{codataold}). The progress in the
determination of the most important fundamental constants for
about 30 years (since the establishment of the CODATA task group)
is shown in Fig.~\ref{f:pro}. The responsibility of the group is
to compare results from different fields and to deliver the most
accurate values of constants important for `precision'
measurements of `essential' quantities. Indeed, the `precision'
threshold is different for different quantities. Note that
constants related to cosmology, astronomy and some from particle
physics such as the Hubble constant, astronomical unit and Cabibbo
angle are traditionally excluded from the consideration as not
being related to the `precision' physics. Meanwhile, certain
properties of light nuclei (deuterium and both stable helium
isotopes $^3$He and $^4$He) are included.

\begin{figure}[hbtp]
\begin{center}
\includegraphics[width=0.6\textwidth]{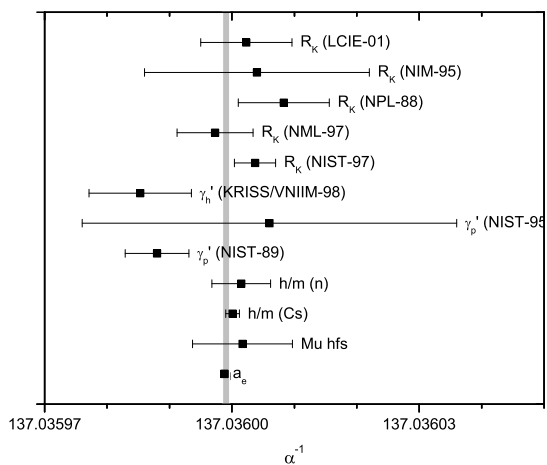}
\end{center}
\caption{Determination of the fine structure constant $\alpha$ by
different methods as discussed in \protect\cite{codata}. Among the
results: a free QED value from the anomalous magnetic moment of
electron ($a_e$), a bound QED value from the muonium hyperfine
structure (Mu), an atomic interferometer value (Cs), a value
involving a lattice parameter (n) and values dealing with
calculable capacitor ($R_K$) and gyromagnetic ratio of proton and
helion ($\gamma$), measured in SI units with help of macroscopic
electric standards. The grey vertical strip is related to the
CODATA-2002 value \protect\cite{codata}. \label{f:alpha}}
\end{figure}

\begin{figure}[hbtp]
\begin{center}
\includegraphics[width=0.6\textwidth]{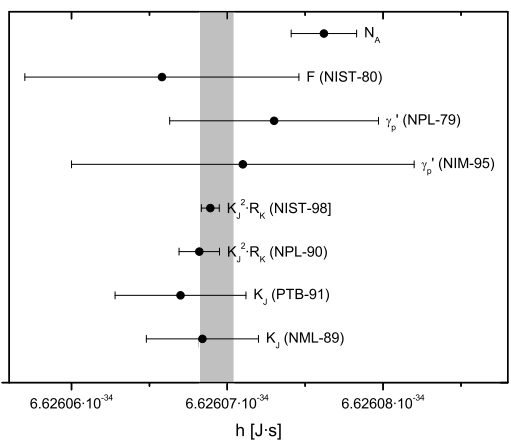}
\end{center}
\caption{Determination of the Planck constant $h$ by different
methods as discussed in \protect\cite{codata}. The results come
from the measurement of the Faraday constants $F=eN_A$, the Avogadro
constant $N_A$, the gyromagnetic ratio of proton ($\gamma_p$),
volt ($K_J$) and watt ($K_J\cdot R_K$) balances. The grey vertical
strip is related to the CODATA-2002 value \protect\cite{codata}.
\label{f:h}}
\end{figure}

The most important lesson we have learned from CODATA's work
is not just their recommended values, but evidence of overall
consistency of our approach to quantitative description of Nature.
That is illustrated in Figs.~\ref{f:alpha} and~\ref{f:h}
showing different approaches to the determination of the fine
structure constant $\alpha$ and the Planck constant $h$
\cite{codata}, which play a central role in the adjustment of the
fundamental constants \cite{codata}.

Why are the values of these two constants so significant for the
CODATA adjustment? The answer is that when we consider physics
from the fundamental point of view, the electron and proton are just
certain particles among many others. However, when we do our
measurements, we deal not with matter in general but mainly with
atomic substances where electrons and protons are fundamental
`bricks'. In such a case electron and proton properties are as
fundamental as $h$ and $c$ or even more. In particular, these
properties determine results of spectroscopy of simple atoms and
macroscopic quantum electromagnetic effects (see
Sect.~\ref{s:gov}). The experiments, results of which are
presented in Figs.~\ref{f:alpha} and~\ref{f:h}, involve, directly
or indirectly, such constants as the Rydberg constant, the
electron and proton masses, the electric charge and magnetic
moments of an electron and a proton, the Planck constant and the
speed of light.

Let us consider the Planck constant in more detail. The
accuracy of the determination of the most important fundamental
constants is summarized in Fig.~\ref{f:pro}. We note, that the
fine structure constant
\begin{equation}
\alpha=\frac{e^2}{4\pi \epsilon_0 \hbar c}
\end{equation}
and the molar Avogadro constant $h\cdot N_A$ have been known
better that some of their constituents, such as $h$, $e$ and
$N_A$. That means that an experimental determination of $h$ is
equivalent to a determination of $e$ and $N_A$ and thus, from the
experimental point of view, the Planck constant $h$ becomes also
related to classical electrodynamics and atomic and molecular
physics of substance. In a `practical' sense, $h$ is even more
universal than in a `fundamental' sense. More discussions on this
subject can be found in \cite{UFN}.

\section{The International System of Units SI:\\ Vacuum
constant $\epsilon_0$, candela, kelvin, mole and other questions\label{s:si}}

From the point of view of fundamental physics, the system SI \cite{SI}
is unnecessarily complicated. It has seven basic units:
\begin{itemize}
\item {\em metre\/}, {\em second\/} and {\em kilogram\/} (which beyond
any doubt are crucial units for any system of units for physical
quantities);
\item {\em ampere\/} (which is already questionable and, in fact,
a number of physicists (see, e.g., \cite{okun}) strongly believe
that it is much better to set $\epsilon_0=1$ and to measure
electrical quantities in terms of three basic mechanical units);
\item {\em kelvin\/} and {\em mole\/} (which are, in a sense,
unnecessary units since the thermodynamic energy and the number of
particles can be measured without introducing any special units);
\item {\em candela\/} (which looks like a worse case --- an
unnecessary unit for a quantity related to a sensitivity of a human
eye, an object, which is rather outside of physics).
\end{itemize}

\subsection{`Unnecessary' units}

\begin{flushright}
\begin{minipage}{9cm}
`I didn't say there was nothing {\em better\/},' the King replied.
`I said there was nothing {\em like\/} it.'\\ \centerline{\em
L.C.}
\end{minipage}
\end{flushright}

Let us start with the `unnecessary' units. From the philosophical
point of view, any measurement is a comparison of two quantities
of the same dimension and thus is a {\em relative\/} measurement.
However, as mentioned above, if we do not like to create every
time a chain of comparisons, we should introduce certain units. A
measurement in terms of these units, although still a comparison, is
a very special comparison and we qualify such as an {\em
absolute\/} measurement. To be more precise, we like to have a
coherent system of units and thus it is not enough to define
any units, we have to define a certain system of units with
\begin{itemize}
\item one unit for each kind of quantity (each dimension);
\item most units derived from a few basic units (e.g., the newton,
a unit of force, is defined through the metre, the second and the
kilogram: $1\; {\rm N} = 1\;{\rm kg} \times 1\;{\rm m} \times
1\;{\rm s}^{-2}$).
\end{itemize}
Meanwhile, in certain areas the relative measurements are so much
more accurate (or much more easier, or have other big advantages)
than the absolute measurements, that we face a hard choice: either
to support a minimized  coherent system of units, or to introduce
some `extra' units (inside or outside the system). There are
several options for a solution. A choice made in the case of
temperature and amount of substance was to extend the system and
to introduce new base units. For the mass of atoms and molecules,
the universal atomic mass unit has been introduced as a unit
outside of SI, but officially recognized and recommended for use.
Nuclear magnetic moments are customarily measured in units of the
nuclear magneton, which has never been included in any official
recommendation of units.

One may think that since the kelvin appeared a long time
ago\footnote{To be more precise, the Celsius temperature scale is
meant since a value of Celsius' and Kelvin's degrees is the same.}
(before we realized that temperature is a kind of energy), it is
kept now for historic reasons only. That is not correct. An
example with the {\em foot\/} in USA shows how to treat the
problem. There is no independent {\em foot\/} there --- this
traditional unit\footnote{Actually there is a number of different
versions of the foot. Eq.~(\ref{foot}) corresponds to the {\em
international foot\/}. There is also the {\em U.S. survey foot\/}
which is equal to ${1200}/{3937}\;{\rm m}$. The number is chosen
in such a way that $1\;{\rm m}=39.37\;{\rm in}$ (see Sect. B.6 in
\cite{SIfoot} for legal detail; historical details can be found in
\cite{SIox}).} is defined as an exactly fixed part of the {\em
metre\/}:
\begin{equation}\label{foot}
1\;{\rm ft} = 0.3048\;{\rm m}~~{\rm (exactly)}\;.
\end{equation}
As a result, for everyday life a use of feet and metres is not
quite the same. Meanwhile, for scientific applications and
industrial precision mechanics and electronics, their simultaneous
use may be quite confusing, but it is completely equivalent: the
same information, the same accuracy, the same actual basic
definitions. On the contrary, the use of the kelvin and the joule
is not the same -- interpreting data from one unit to the other
changes the accuracy of the results.

\subsection{`Human-related' units}

\begin{flushright}
\begin{minipage}{7.5cm}
It was labelled `Orange marmalade', but to her great disappointment
it was empty.\\ \centerline{\em L.C.}
\end{minipage}
\end{flushright}

The case of {\em candela\/} presents an additional problem, which is
not a question of units, but a question of quantities. Why did the
{\em original\/} foot and similar units fail? One of the reasons
is that they were ill-defined. But that is half-truth only.
The truth is that they were
related to non-physical quantities. The original foot was first
related to a size of a particular person (a king/queen), later some
approaches were related to an `average' person. And only
eventually, the `human' foot was substituted by an artificial
ruler.

When we rely on properties of a particular or average human being,
we deal with a biological object. If we now reverse the problem
and try to check a value related to the foot in its
original sense, we meet a biological problem. We need to make a
decision on the selection of people, to address the problem that
the result may depend on geography etc. Eventually, the ultimate
decision will choose either a `conventional foot' (as it is),
which is not related anymore to any person, or a `conventional
person', which should be a subject of a real measurement. In other
words, even measuring some property in well-defined units, we may
need in certain cases an {\em arbitrary\/} agreement on what
this property is. We qualify this kind of agreement as {\em
arbitrary\/} because within certain margins we are free to adopt
any parameters.

In fact, the problem of the average size of a human body is not so
important now, however, there is a number of questions due to
ecology, safety, medicine, which involve interactions of
certain physical effects and a human being. We can easily
characterize these phenomena by a complete description of their
physical properties. However, for obvious practical reasons, we
often need an {\em integral\/} estimation of the influence which
involves a number of parameters, which values vary in a broad
range (such as frequency), and we certainly know that the human
sensitivity depends on frequency and other various parameters.
Such integral characteristics are not of physical nature. To
determine them we need to perform two kinds of measurements on
\begin{itemize}
\item physical details of the effects (a kind of the spectral
distribution);
\item spectral sensitivity of a human being.
\end{itemize}

If we accept the sensitivity as a real quantity, which is
determined by effects beyond physics, the whole integral
characteristic becomes not a pure physical but of combined nature:
physics$+$biology. If we accept a model for the sensitivity, we
can do simple calculations within this model and obtain a pure
physical result, which will be related to the model rather than to
reality. In other words, the result will be in well-defined units
but for a conventional quantity. In some cases (e.g., in radiology
\cite{icru}) real and conventional quantities are clearly
distinguished. In others the separation is less clear. But in any
case, quantities, related to a {\em sensitivity\/} of an [{\em
average}] human eye, cannot be accepted as physical quantities and
it is does not matter how their units are defined.

What is also important is a status of the SI system as an
international treaty. Everything related to the SI is a part of
this agreement. Otherwise, it is not a part of the SI. A unit has
to be a unit for a certain quantity. If a quantity is not
well-defined, the unit is also ill-defined. If we deal with a
quantity for which an additional agreement is needed, we have to
put it into the SI, because it has to be a part of a definition of
the related unit.

The question of the candela is very doubtful. The candela itself
is defined as a part of the SI in `rigid physical
terms'\footnote{{\em The candela is the luminous intensity, in a
given direction, of a source that emits monochromatic radiation of
frequency $540\times10^{12}$ hertz and that has a radian intensity
in that direction of 1/683 watt per radian\/} \cite{SI}.}. As we
mentioned above, there may be a need to have a convention on
properties, but never on all physical quantities of a kind. We
define the metre of the SI and we can use it. Length in general is
well-defined and does not need any additional agreement. However,
if we like to measure particular properties of certain objects,
which are related to length, we may need an additional agreement
on these properties. For example, when we deal with an average
parameter of a human body.

This problem of a `conventional' characteristic or a
`conventional' object is not only for human-related (or
life-related) cases. It is due to the peculiarity of classical
objects. A number of well-known not-life-related examples of
conventional properties are related to those of Earth such as the
`standard acceleration due to gravity'
($g_n=9.806\,65\;{\mbox{m/s}}^2$), adopted by General Conference
on Weights and Measures, the `standard pressure' (of $1\;{\rm
atm}=101\,325\times10^5\;{\rm Pa}$), various conventional days and
years. From a formal point of view, these values have nothing to
do with the real acceleration of free fall which varies from place
to place and during the day. As we mentioned in Sect~\ref{s:art},
the metric treaty first relied to properties of Earth (the metre,
the second, and, indirectly, the kilogram), which were believed to
be well-defined. Later, it was realized that they are not and the
units were redefined via artificial objects (which are now partly
substituted by natural quantum objects). Earth as a whole also
presents an example of a conventional object, when its shape is
simplified and a number of properties are `projected' to the sea
level. A convention is not necessary related to a peculiar object,
a reason for a convention may be specific conditions of
experiment. A recommendation for the practical realization of the
metre \cite{meter} gives a list of accurately measured optical
atomic and molecular transitions. However, some transitions are to
be measured under specific conditions, i.e. their given
frequencies are {\em related\/} to real transition frequencies
{\em not\/} necessarily identical to them.

The case of the candela and photometry is very specific and quite
different from any other basic unit. Without accepting, as a part
of the SI, a convention either on a spectral sensitivity of a
human eye\footnote{Appendix 2 of the official SI booklet \cite{SI}
contains some details of practical realizations of all basic SI
units. In the case of the candela, it reads: {\em The definition
of the candela given on page 98\/} [of \cite{SI}] {\em is
expressed in strictly physical terms. The objective of photometry,
however, is to measure light in such a way that the result of the
measurement correlates closely with the visual sensation
experienced by a human observer of the same radiation. For this
purpose, the International Commission on Illumination (CIE)
introduced two spectral functions $V(\lambda)$ and
$V^\prime(\lambda)$...} One of them, $V(\lambda)$, is applied in
photometry. However, the recommendations of CIPM on practical
realizations do never (except of the case of the candela) contain
any information which in {\em needed\/} for the realization. They
are supposed to deliver certain information which follows from the
main body of the SI booklet and {\em may\/} be used to simplify
the realization (see, e.g., Sect.~\ref{ss:macro} and
\ref{s:mise}).} or on what are `the same sensation' and `an
[average] human eye' we see a very reduced field for measurements.
Actually, there are two kinds of photometrical quantities: visual
and physical. To introduce the candela as a unit for both, we need
both kinds of conventions.

The candela and photometrical quantities were designed to deal
with all visible frequencies. However, at the present time the SI
system does not include any convention which allows to go beyond
the frequency of $5.4\times10^{14}\;{\rm Hz}$ at which the candela
is defined. That means that the candela definition as an SI unit
is incomplete and completely compromises it as such, because the
SI denies any quantity to be measured in candelas. Within just the
SI, we cannot measure photometrical quantities related to, e.g.,
red light.

For really physical quantities (such as electrical current or
amount of substance) those definitions are rigid, however, for the
human-related quantities, the definitions are quite flexible and
need additional assumptions to be adopted. There may be different
opinions on what the best way to treat the candela is and how to
modify the SI for that. However, we have to acknowledge that in
the current version of the SI system \cite{SI} the candela, as an
SI unit is much compromised and cannot be barely used as an SI
unit for any application. The physical quantities are defined
through physical laws and that means that they are `defined by
Nature', not by us.

\subsection{Vacuum constant $\epsilon_0$ and Gaussian units}

\begin{flushright}
\begin{minipage}{8.5cm}
`You can call it ``nonsense'' if you like', she said, `but I'{\em
ve\/} heard nonsense, compared with which that would be as
sensible\- as dictionary!'\\ \centerline{\em L.C.}
\end{minipage}
\end{flushright}

Although the candela is the most questionable among the basic SI
units, it has never been a subject of a world-wide discussion as
the ampere and the Gaussian units have been. Obviously, that is
because of the significance of electromagnetic phenomena in modern
physics. There is no doubt that the Gaussian units, in which
$\epsilon_0=1$, are better for understanding of electrodynamics.
However, there is a number of units very well suited for some
classes of phenomena (see, e.g., Sect.~\ref{s:unit}) and that does
not mean that these units are proper units for general purposes.
In this short chapter I try to explain why the units with
$\epsilon_0=1$ have never been good for the general use.

First, we have to remind that the units are needed mainly
to express results of measurements (done or predicted). If these
practical units are not good for theory, we may do calculations in
more appropriate units, but at the end we have to present the
final results in some practical units.

Secondly, we remark that there are some areas and, in particular,
a field of electrotechnical measurements (of electric potential,
current, resistance, inductance and capacity), where relative
measurements can be done much simpler and more accurately (in
respect to absolute measurements of the same values). Why is it
so? The answer is simple: both the SI and Gaussian definitions of
the basic electromagnetic units involve calculations of the
magnetic or electric fields and building of a macroscopic bulk
setup with well controlled  values of these fields. In other
words, the absolute measurements deal with a completely different
kind of experiments. The absolute measurements correspond to
electrodynamics, while the relative measurements of quantities
listed above are related to electrotechnics.

For this reason, nearly all electric measurements are realized as
relative measurements done in special `electrotechnical' units.
Separate experiments are performed in a limited number of
metrological laboratories to cross-check these units and to
calibrate them properly in terms of the SI. At earlier times, the
standards were built on classical objects. They were artificial
and in this sense similar to the present standard of mass.
However, in contrast to a prototype of the kilogram, they were
much more vulnerable. There is a number of effects which may
affect properties of classical objects and shift them. However, it
is much easier to `break' an electric device than a weight. Thus,
the electrical units evolved and their calibration was not a
simple procedure. They were quasi-independent. That produced a
strong need to have an independent unit for electrical effects and
to provide it one has to have $\epsilon_0\neq1$. A value of
$\epsilon_0$ has been fixed within the SI, but it was unknown in
practical units and had to be measured.

Now, we have taken an advantage of the application of macroscopic
quantum effects (see Sect.~\ref{ss:macro} and \ref{s:mise} for
detail) and may be sure that the practical electrical units do not
evolve, but still we need to calibrate them. While a value of
$\epsilon_0$ is calculable in terms of SI units
\begin{eqnarray}
\epsilon_0 &=& \frac{1}{c^2\mu_0}\nonumber\\
&=&\frac{10^7}{4\pi\bigl(299\,792\,458\bigr)^2} \;\mbox{F/m}\nonumber\\
&=& 8.854\;187\;817...\times 10^{-12}\;\mbox{F/m}\,
\end{eqnarray}
it is still unknown in practical units (such as, e.g., ohm-1990,
$\Omega_{90}$, \cite{Ohm901,Ohm902}) and is a subject of
measurement. Likely in future we will decide to reverse a
situation accepting quantum definitions of ohm and volt. That will
upgrade today's practical units $\Omega_{90}$ \cite{Ohm901,Ohm902}
and $V_{90}$ \cite{Volt90} up to the status of SI units, but will
make $\epsilon_0$ a measurable quantity and will substitute the
prototype of the kilogram by an electrical balance. In such a
scenario the values of the Planck constant $h$ and the elementary
charge $e$ would be fixed and, with a value of the speed of light
already fixed, one can see that
\begin{equation}
\epsilon_0 =\frac{e^2}{4\pi \alpha  \hbar c}\;,
\end{equation}
where the fine structure constant $\alpha$ as a dimensionless
constant has an unknown value, which is a subject of measurement.
Thus, $\epsilon_0$ becomes a measurable quantity certainly not
equal to unity in any sense. That is why we should not like to set
a simple identity $\epsilon_0=1$ now.

There is one more issue about the constants of vacuum $\epsilon_0$
and the fine structure constant $\alpha$. We may wonder whether
the fine structure constant is calculable or not. We cannot answer
this question now, however, there is a certain constraint on a
scheme how $\alpha$ might be predicted. The most expected scenario
is that we would be able to predict $\alpha_0$ (a value of the
fine structure constant related to the Planck scale) as a kind of
a geometric factor (see Sect.~\ref{s:calc}, \ref{s:geo} and
\ref{s:planck}). In such a case, the electric charge would not be
a new independent property, but a kind of a derivative from
mechanical properties and should be, in principle, measured in
mechanical units with $\epsilon_0=1$. However, if $\alpha$ has a
value chosen due to spontaneous breakdown of symmetry (see
Sect.~\ref{s:geo} and \ref{s:const}) it may be any.
Perhaps, we should treat the electric charge as a new independent
quantity and measure it in separate units with a dimensional value
of $\epsilon_0$ as a consequence. A situation with the Coulomb (or
Ampere) law is different from, e.g., that with the Newtonian
gravity. For the latter we have already known that the gravitational
charge (i.e., the gravitational mass), due to a deep physical reason
(the equivalence principle), is not a new property, but a derived
property completely determined by the inertial mass. Indeed, the
question of calculability of $\alpha$ and thus of the origin of the
electric charge will not be answered soon and, indeed, such a
general view would never affect any decision on units, because of
practical importance of the question. The practice of electrical
measurements has obviously pointed to the proper choice.

\subsection{`Unnecessary' units, II}

While discussing `necessary' and `unnecessary' units, we would like to
mention a point important for practical use. When we speak about
most of phenomena, we often apply a `jargon' dropping important
words. In such a case to understand anything properly,
redundant information would be helpful. For example, we often speak
about a magnetic field not clearly discerning magnetic induction $B$ and
magnetic field strength $H$ (which are not simply related in
media), or even about a field not specifying whether we mean
magnetic or electric field. In such a case, the use of different
units is very helpful to understand the practical situation. The
same story is with units and their biological equivalents, which
from a theoretical point of view should be rather the same. However,
naming the unit we immediately explain which property of, e.g.,
radiation we have in mind: their energy or their effect on a human
body. A choice of a unit plays a role of a flag allowing to drop a
number of words. Use of four different units for the electromagnetic
field (for $E$, $D$, $B$ and $H$) makes theory less transparent
and unnecessary complicated, however, these four units may be
helpful to describe an experiment in a much shorter way.

\section{Physical Phenomena Governed by Fundamental Constants\label{s:gov}}

\begin{flushright}
\begin{minipage}{6cm}
It's as large as life and twice as natural! \centerline{\em L.C.}
\end{minipage}
\end{flushright}

We mentioned above a quantum approach to standards of electrical
units. They superseded classical standards providing universal
values which do not depend upon time or location of the
measurement\footnote{We discuss possible time- and space
variations of fundamental constants in Sect~\ref{s:const} and
\ref{s:search}.}. It is also very useful that we can determine
them from certain experiments not related to electricity. That is
because they are based on fundamental constants. However,
fundamental constants, if they are really fundamental in some
sense, should show themselves only at a fundamental level, while
any particular measurements deal with objects and phenomena
far from fundamental. How can we access any fundamental
quantity? The obvious answer is that we have to try to find a
property of a certain non-fundamental object, which we can
calculate. There are two general kinds of such objects.
\begin{itemize}
\item First, we can study relatively simple objects, which properties
can be calculated by us. The simplest are
particles, and only recently we learned how to study a single
particle in a trap. In earlier times we dealt with beams and
clouds of interacting particles trying to eliminate their
interactions. The next in the row of simple objects are simple
atoms and simple molecules.
\item Another option are macroscopic quantum effects, such as, e.g.,
the Josephson effect. Once we realized proper conditions, we can
see the same result for various samples and the result is simply
expressed in terms of fundamental constants. An important feature
of this kind of effects is that when conditions are not perfect,
they often make the effect harder or even impossible to observe,
but seldom affect the basic parameters of the effect. So, certain
quantities coming from this kind of effects are quite immune to
the conditions of the experiment.
\end{itemize}

Before we discuss any application of properties of elementary and
compound particles, let us underline that an important detail for
the interpretation of such measurements is that particle
properties are the same for each species. A measurement may be
even of classical nature aiming to determine the Avogadro or
Faraday constant, however, the output is very different in
classical and quantum framework. For example, the Faraday constant
\begin{equation}
F = e\, N_A
\end{equation}
in the classical consideration is defined for an {\em average\/}
charge carried by the Avogadro number of electrons (correcting for
the sign of the charge) or single-charged ions, while $N_A$ is in
turn an average number of carbon atoms needed to form 12 gram of a
carbon material. In the quantum case, we know that carbon atoms or
electrons are the same and we can drop a word `average'. In the
quantum case the electron charge $e$ is certainly a fundamental
constant, while from a classical point of view it is about the
same as an average mass of dust particles or rain drops.

Classical physics is unable to deal with identical objects. What
does `identical' mean? If there is no interference, we can always
distinguish between two electrons. From the point of view of
classical statistics `identical' means `different recognizable
objects for which in a particular consideration we do not care
which is which\footnote{Historically, the statistical analysis
appeared in an attempt to describe social phenomena dealing with
people. Definitely, that is just a case when the objects are
clearly distinguishable.}'. But if we did care, we could always
recognize them. If two electrons have approximately the same mass
and charge, classical mechanics cannot check if they are the same
exactly or approximately, because there is always an experimental
uncertainty. It may be in principle reduced to any level but never
removed completely.

Quantum physics introduces interference between particles. Physics of two
slightly different electrons and two identical electrons is by far
not the same. And what is very important, we do not need to
perform any interferometric experiments. They have already been
performed for us by Nature. The Pauli principle governs the atomic
levels and nuclear shells. All electrons are identical as well as all
protons and neutrons. The very existence of lasers showed the
identity of all photons as particles.

Identity of objects of the same class makes their properties to be
natural constants. However, if we like them to be really
fundamental from a theoretical point of view, we have to study
simple objects.

\subsection{Free particles}

Studying free particles offers relatively limited access to the
fundamental constants. We can measure their masses and magnetic
moments. Their electric dipole moments have been searched for (because
of different supersymmetrical theories) but have not yet been detected.
Their charge is known in relative units and seems to be trivial.
Sometimes, but very seldom, they have calculable properties. Two of the
most important of them are related to the anomalous magnetic moment of
electron and muon.

Briefly speaking, if we like to learn something beyond basic
properties (such as mass) of an object, we have to study
interactions. An interaction with a classical field is not a good
case because we can hardly provide configurations of the classical
electromagnetic field controlled with a high accuracy. Only one
kind of classical fields is suited for precision measurements,
namely, a homogeneous magnetic field with a fixed, but
unknown (in the units of SI) strength. That allows us to compare
masses and magnetic moments. A quantum interaction is under much
better control, because its strength is controlled by Nature and not
by us. Measuring the mass or the magnetic moment, one determines
certain fundamental parameters directly, while dealing with a
calculable interaction and calculable properties, we access
fundamental constants indirectly. A quantum electrodynamical
self-interaction allows us to present the anomalous magnetic
moment of an electron $a_e$ in terms of the fine structure
constant with a high accuracy \cite{aeth}
\begin{eqnarray}
a_e &=& \frac{1}{2}\frac{\alpha}{\pi}-0.328\,478\,696
\left(\frac{\alpha}{\pi}\right)^2 + 1.181\,241
\left(\frac{\alpha}{\pi}\right)^3 - 1.737(39)
\left(\frac{\alpha}{\pi}\right)^4+\dots\nonumber\\ &+& {\it
near~negligible~effects~of~weak~and~strong~interactions}\;.
\end{eqnarray}
A measurement of $a_e$ \cite{aeexp}
\begin{equation}
a_e = 1.159\,652\,188\,3(42)\times10^{-3}~~~~[3.2\times10^{-9}]
\end{equation}
delivered the most accurate value of $\alpha$
\begin{equation}
\alpha^{-1}_{\rm
g\!-\!2}=137.035\,998\,80(52)~~~~[3.4\times10^{-9}]\;.
\end{equation}
In Table~\ref{t:part} we list fundamental constants which may be
obtained studying particles. An example of a non-elementary
particle is the deuteron. Measuring its binding energy and masses
of proton and deuteron, one can obtain the neutron mass
(see Sect.~\ref{ss:comp}).

\begin{table}[hbtp]
\begin{center}
\begin{tabular}{c|c|c}
  \hline
Particle & Constant & Comment \\
 \hline
Electron & $m_e$ &  via comparison to proton\\
 & $\alpha$ &  via QED\\
Muon &$a_\mu$  & via comparison to proton\\
 &$\mu_\mu/\mu_p$  & via comparison to proton\\
Neutron & $\alpha$ &  via comparison to lattice spacing and $R_\infty$\\
Deuteron & $m_n$ &  via a measurement of binding energy and comparison to proton \\
Caesium & $\alpha$ &  via atomic interferometry and Raman scattering\\
 \hline
\end{tabular}
\end{center}
\caption{Fundamental constants determined through properties of
elementary and compound particles\label{t:part}}
\end{table}

\subsection{Simple atoms and molecules\label{S:simple}}

From the very beginning of studies of classical effects, we
distinguished kinematics (i.e., a theory of particle motion at a
given force or a given potential field) and dynamics (i.e., a
theory of forces: fields, their sources and their interactions
with particles). Classical theory of Maxwell equations is a
dynamics of charged particles and kinematics of photons. Quantum
mechanics introduced quantized properties and identical objects.
That, additionally to dynamics and kinematics, opens an
opportunity for a prediction of structure of certain objects.
Indeed, classical physics could successfully consider compound
objects, which consist of `simple' constituents such as the Solar
system formed by the Sun and the planets. However, there has been
no chance for any {\em ab initio\/} calculation, because physics
deals only with particular objects and in classical physics any
particular object is a peculiar object: the parameters are
peculiar and the initial conditions are peculiar. Quantum physics
introduced fundamental particles: a few kinds of particles which
form everything in our World. We still need to know some their
parameters, but with these few parameters determined we can try to
calculate everything. We can also reverse the problem: predicting
the structure properties in terms of fundamental parameters of
constituent particles and determining the properties
experimentally, we deduce actual values of the parameters.

If we study a few-particle system, we have a better chance to study
interactions of its constituents. However, different simple atoms
are treated within QED in very different ways. The most clear case
is the lepton quantum electrodynamics (electron and muon are
the most important leptons in QED). It needs very few input data:
the elementary charge and the lepton masses. Any other property
(e.g., the magnetic moment) can be in principle derived
from these few.

Indeed, the proton cannot be treated this way. Generally speaking,
QED is a theory of electromagnetic interactions of leptons, photons
and `external' sources. A proton is such a source and we need a
number of parameters to describe it: its charge $q_p$, magnetic
moment $\mu_p$, details of its charge and magnetic moment
distribution as well as more sophisticated parameters. What we only know
is that this approach is consistent
--- we can measure these parameters and they are the same for very
different kinds of experiments.

Theory of simple atoms involves a few dimensionless parameters
(which actually are small parameters for a number of important
applications and allow various perturbative expansions):
\begin{itemize}
\item the fine structure constant $\alpha$, the exponential factor of
which indicates how many QED loops are taken into account;
\item the Coulomb strength $Z\alpha$, with the nuclear charge $Z$
changing in a very broad range from $Z=1$ to $Z\simeq 90$;
\item the mass ratio of the orbiting particle to the nucleus, which is
$m_e/Am_p\simeq 10^{-3}/2A$ ($A$ is the nuclear mass number) for
a conventional (`electronic') atom; $m_e/m_\mu\simeq 1/207$ for
muonium (the nucleus is a positive muon); unity for positronium
(the nucleus is a positron); and
$m_\mu/Am_p\simeq 1/9A$ for a muonic atom;
\item various parameters related to the nuclear structure.
\end{itemize}
It is clear that an exact calculation with all these parameters is
not possible. Some calculations for conventional atoms are exact
in $Z\alpha$, while positronium calculations apparently must be
done exactly in the electron-to-nucleus mass ratio. Still, an
expansion in other small parameters has to be applied and an
accurate theory is possible not for any simple atoms. A proper
estimation of uncalculated terms is sometimes a difficult problem
\cite{PSASqed}. Various details of theoretical calculations can be
found in review \cite{report}.

In light atoms, the perturbative approach is dominant and to
demonstrate how far we can go with theoretical predictions, we
summarize in Table~\ref{t:order} crucial (for a comparison with
experiment) orders of QED theory for energy levels in various
two-body atoms. We remind that the leading non-relativistic
binding energy is of order of $(Z\alpha)^2m_ec^2$.

\begin{table}[hbtp]
\begin{center}
\begin{tabular}{lc}
\hline
Value & Order \\
 & [in units of $(Z \alpha)^2m_ec^2$] \\
\hline
Hydrogen, deuterium (gross structure) & $\alpha(Z\alpha)^5$, $\alpha^2(Z\alpha)^4$ \\
Hydrogen, deuterium (fine structure)              & $\alpha(Z\alpha)^5$, $\alpha^2(Z\alpha)^4$ \\
Hydrogen, deuterium (Lamb shift)              & $\alpha(Z\alpha)^5$, $\alpha^2(Z\alpha)^4$ \\
$^3$He$^+$ ion ($2s$ HFS)  & $\alpha(Z\alpha)^5m_e/M$, $\alpha(Z\alpha)^4m_e^2/M^2$,\\
& $\alpha^2(Z\alpha)^4m_e/M$, $(Z\alpha)^5m_e^2/M^2$\\
$^4$He$^+$ ion (Lamb shift)              & $\alpha(Z\alpha)^5$, $\alpha^2(Z\alpha)^4$ \\
N$^{6+}$ ion (fine structure)              & $\alpha(Z\alpha)^5$, $\alpha^2(Z\alpha)^4$\\
Muonium ($1s$ HFS)      & $(Z\alpha)^5m_e^2/M^2$, $\alpha(Z\alpha)^4m_e^2/M^2$,\\
&$\alpha(Z\alpha)^5m_e/M$ \\
Positronium ($1s$ HFS)  & $\alpha^5$ \\
Positronium (gross structure)      & $\alpha^5$ \\
Positronium (fine structure)       & $\alpha^5$ \\
\hline
\end{tabular}
\end{center}
\caption{Crucial (for a comparison of QED theory and experiment)
orders of magnitude for corrections to the energy levels in units
of $(Z \alpha)^2m_ec^2$ (see \cite{PSASqed} for detail). Here: $M$
stands for the nuclear mass.\label{t:order}}
\end{table}

Various simple atoms and certain simple molecules can deliver us
much more information on fundamental constants than free
particles, because we are able to express their properties in
terms of such fundamental constants as the Rydberg constant
$R_\infty$, the fine structure constant $\alpha$, various masses
($m_e$, $m_p$, $m_\mu$, $m_\pi$ etc.), magnetic moments ($\mu_p$,
$\mu_d$, $\mu_\mu$ etc.) and some other constants. Working with
atoms and molecules we can apply various spectroscopic methods,
which are the most accurate at the moment.

Simple molecules are much more complicated than simple atoms and
their use is rather limited. For example, studies of the hydrogen
deuteride (HD) provide us with the most accurate value of
$\mu_p/\mu_d$ \cite{plamudp}. A summary on use of simple atoms and
molecules to determine precision values of various fundamental constants is given in
Table~\ref{t:atmol}. More details on simple atoms can be found in
\cite{psas2000,psas2002}, while a popular history of applications
of hydrogen to fundamental problems is presented in \cite{rigden}.

\begin{table}[hbtp]
\begin{center}
\begin{tabular}{l|c|c}
  \hline
System & Constant & Comment \\
 \hline
Muonium & $\alpha$ &  via bound state QED\\
 & $m_\mu/m_e$ &  via bound state QED\\
 & $\mu_\mu/\mu_p$ &  via bound state QED and comparison to proton\\
Hydrogen & $R_\infty$ & via bound state QED\\
 & $\mu_p/\mu_e$ &  via bound state QED\\
Deuterium  & $R_\infty$  &  via bound state QED\\
  & $\mu_d/\mu_e$  &  via bound state QED\\
Helium & $\alpha$  &  via bound state QED\\
Hydrogen-like carbon & $m_e/m_p$ &  via bound state QED\\
Hydrogen-like oxygen & $m_e/m_p$ &  via bound state QED\\
Muonic atoms & $m_\mu/m_e$ &  via bound state QED\\
Pionic atoms & $m_\pi/m_e$ &  via bound state QED\\
HD molecule & $\mu_d/\mu_p$ &  via bound state QED\\
HT molecule & $\mu_t/\mu_p$ &  via bound state QED\\
 \hline
\end{tabular}
\end{center}
\caption{Fundamental constants determined through simple atomic
and molecular systems\label{t:atmol}}
\end{table}

\subsection{Free compound particles\label{ss:comp}}

Free particles, which we can study, are not necessarily elementary
particles. We can treat nuclei, atoms and molecules as compound
particles and study their simplest properties (such as the mass or
the magnetic moment). We can also rely on conservation laws. For example, the
best value of neutron mass comes from deuteron studies. The deuteron
mass \cite{ame}
\begin{equation}
m_d = 2.013\,553\,212\,70(35)\; \mbox{\rm
u}~~~~[1.7\times10^{-10}]
\end{equation}
combined with an accurate value of its binding energy $E_d$ of
approximately 2.2 Mev \cite{edeu} and the proton mass \cite{mpro}
\begin{equation}
m_p = 1.007\,276\,466\,89(14)\; \mbox{\rm
u}~~~~[1.4\times10^{-10}]\;,
\end{equation}
provides us with such a possibility
\cite{ame,codata}\footnote{This value of the neutron mass also
involves data similarly related to other, more complicated, nuclei
\cite{ame}. Note that the data on the binding energy may differ
from the originally published results because of the recalibration
of the lattice parameter.}
\begin{eqnarray}
m_n &=& m_d - m_p - E_d/c^2\nonumber\\
&=&1.008\,664\,915\,60(55)\; \mbox{\rm u}~~[5.5\times10^{-10}]\;.
\end{eqnarray}

Another example of an application of compound particles:
determination of the fine structure constant $\alpha$ via
scattering of photons at the caesium atom and measuring their
recoil shift \cite{chu}. In this experiment, one should deal with
absorption and stimulated emission, however, dynamical details of
the interaction of the photons and the atom are unimportant. Once
we know the direction and frequency of the photons, we need to
know only about the atom as a whole: its energy and momentum. If
we treat atom in such a way as a compound particle, it is `simple'
in a sense.

We remind that properties of certain atoms and molecules play a
crucial role in the definition of SI units: the hertz is defined
via the hyperfine interval in the caesium-133 atom and the kelvin
via the triple point of the water. The universal atomic mass unit
(a non-SI unit acceptable along with the SI) and the mole are
defined via the mass of the carbon-12 atom. However, considering
atomic properties as units, we indeed should not care if they may
be calculated.

\subsection{Macroscopic quantum phenomena\label{ss:macro}}

Macroscopic quantum phenomena offer us certain properties which
may be presented in terms of fundamental quantities. For example,
the Meissner effect provides us with a quantized magnetic field.
The magnetic flux through a superconducting loop can take only
very specific values such as
\begin{equation}
\Phi_n = n \Phi_0\,,
\end{equation}
where
\begin{equation}
\Phi_0 = \frac{h}{2e}
\end{equation}
is the magnetic flux quantum and $n$ is an integer number. As we
mentioned, the most important for applications is to consider
macroscopic quantum phenomena which are related to quantized
values of electrotechnical properties. Two such effects play
crucial role for practical use. These are the quantum Hall effect
and the Josephson effect. The former offers us a quantized value
of the resistance
\begin{equation}
R_n=\frac{R_K}{n}\;,
\end{equation}
proportional to the von Klitzing
constant
\begin{equation}
R_K = \frac{h}{e^2}\;,
\end{equation}
while the latter allows to quantize the voltage related to some
frequency $\nu$:
\begin{equation}
U_{n^\prime} = \frac{n^\prime}{K_J} \cdot \nu\;,
\end{equation}
where
\begin{equation}
K_J = \frac{2e}{h}
\end{equation}
is the Josephson constant and $n$ and ${n^\prime}$ are certain
integer numbers.

We note that a direct measurement of these constants in SI units
is very complicated, and consider different approaches to the
application of these two quantized phenomena in
Sect.~\ref{s:mise}.

\subsection{Atomistics and discrete classical phenomena}

As already mentioned, even certain classical constants such as the
Avogadro and Faraday constant have quantum origin. First, they
originate from the atomic nature of substance, which is a consequence of
quantum mechanics, and, secondly, they receive their meaning
because of identity of species of the same kind.

We collect in Table~\ref{t:codata} the most accurately known
fundamental constants \cite{codata} which have been obtained
through studies of calculable objects.

\begin{table}[hbtp]
\begin{center}
\begin{tabular}{c|c|c}
  \hline
Constant & CODATA 2002 values & Method \\
 \hline
$R_\infty$     &  $10\,973\,731.568\,525(73)\; \mbox{\rm m}^{-1}~~~~[6.6\times10^{-12}]$ & bound state QED \\
$m_p$ & $ 1.007\,276\,466\,88(13)\; \mbox{\rm u}~~~~[1.3\times10^{-10}]$& free particles \\
$m_p/m_e$      &  $1\,836.152\,672\,61(85)~~[4.6\times10^{-10}]$& bound state QED, free particles \\
$a_e$ & $1.159\,652\,185\,9(38)\times10^{-3}~~~~[3.2\times10^{-9}]$& QED, free particles \\
$\alpha^{-1}$       &  $137.035\,999\,11(46)~~~~[3.3\times10^{-9}]$& QED, free particles, bound state QED, MQE \\
$\mu_p/\mu_B$  & $1.521\;032\,206(15)\times10^{-3}~~[1.0\times10^{-8}]$& bound state QED \\
$m_\mu/m_e$    & $206.768\,283\,8(54)~~~~[2.6\times10^{-8}]$& bound state QED \\
$\mu_\mu/\mu_p$ & $- 3.183\,345\,118(89)~~~~[2.6\times10^{-8}]$& bound state QED \\
$e$ & $1.602\,176\,53(14)\times10^{-19}\;{\rm C}~~~~[8.5\times10^{-8}]$& MQE, ADN \\
$F$ & $96.485\,383\,3(83)\times10^{23}\;{\rm C/mol}~~~~[8.6\times10^{-8}]$& MQE, ADN \\
$\mu_p$ & $1.410\;606\;71(12)\times10^{-26}\;{\rm J/T}~~~~[8.7\times10^{-8}]$&free particles, MQE, ADN\\
$h$ & $6.626\,069\,3(11)\times10^{-34}\;{\rm J\,s}~~~~[1.7\times10^{-7}]$& MQE, ADN \\
$N_A$ & $6.022\,141\,5(10)\times10^{23}\;{\rm mol}^{-1}~~~~[1.7\times10^{-7}]$& MQE, ADN \\
$a_\mu$ & $1.165\,919\,81(62)\times10^{-3}~~~~[5.3\times10^{-7}]$& free particles, QED \\
\hline
\end{tabular}
\end{center}
\caption{CODATA 2002 recommended values of some fundamental
constants \cite{codata} and methods applied to achieve their
values. Here: MQE stands for macroscopic quantum effects, ADN for
effects due to atomic discrete nature of
substance\label{t:codata}}
\end{table}

\section{Fundamental Constants and Renormalization:\\ Operational philosophy of physics\label{s:phil}}

\begin{flushright}
\begin{minipage}{8cm}
And she tried to fancy what the flame of a candle is like after
the candle is blown out, for she could not remember ever having
seen such a thing.\\ \centerline{\em L.C.}
\end{minipage}
\end{flushright}

Quantum mechanics appeared after a brilliant success of relativity
and indeed it was understood that a non-relativistic quantum
theory should be extended to the relativistic case. However, the
development met a problem that perturbative calculations have
involved certain divergencies. The problem has been solved by the
introduction of a procedure of renormalization.

The solution of this problem has a philosophical side and before
addressing the problem, let us discuss some philosophical aspects
of physics and, first of all, answer a question what objectives of
physics are. Let us do that pragmatically. We do not discuss what
various sciences pretend to aim, we like to check what they really
do. Both philosophy and physics pretend to understand Nature.
Philosophy picks up the most significant questions of the very
existence of Nature, however, it does not care if we have enough
data to answer them. And actually, similar to the truly
fundamental constants, the fundamentality never shows itself for
measurements. Physics also pretends to understand Nature, however,
in reality it does not care what Nature, matter or any particular
object such as a photon and an electron are. Physics questions not
what various objects are, but how they interact to each other. It
studies not what Nature is, but how it operates. We, physicists,
certainly believe that something really exists in an `absolute'
sense since the same experiments produce the same results.
However, we cannot say that anything particular exists until we
measure it. It is close to the positivistic philosophy. However,
that is not a philosophy of physicists, but a kind of {\em modus
operandi\/} in physics. This kind of double standards is often met
in everyday personal and professional life: there is a philosophy,
which provides us with a general view on events and there is an
operational scheme, which determines our reaction to the events.
The philosophical views of physicists on Nature may be very
different from each other, while their professional operational
scheme is nearly the same for anybody. This scheme is based on a
kind of a `short-range' philosophy. I call the philosophy beyond
the operational scheme as `operational philosophy'.

Indeed, we may say that matter exists. Or that an electron exists.
But that gives us no real piece of information at all. If we could say
that a certain particle with specific properties exists, that would contain
certain information, and could be correct or not. But to learn
that we have to perform an experiment.

A philosophical breakthrough of special relativity was the idea
that the simultaneousness of events are unmeasurable. Quantum
mechanics said that trajectory is unmeasurable. That we cannot
distinguish between two identical particles. That we cannot
measure certain properties simultaneously. That we cannot do
`exact' measurements without certain consequences. After we had
learned that, we changed our view on what exists and what does
not.

Non-relativistic quantum mechanics succeeded with a perturbative
approach. We start with an unperturbed equation with unperturbed
parameters and introduce various small perturbations which shift
properties of the result. In quantum mechanics we are able `to
switch off' most of perturbations for real quantum mechanical
problems or at least vary their parameters. On the contrary, in
quantum electrodynamics (QED) we cannot turn off the
self-interaction. It is proportional to a small parameter
$\alpha\sim 1/137$, but because of the divergencies the
perturbative correction is not small. It cannot be even calculated
properly because it involves physics of high momenta. QED said,
that since the unperturbed `bare' parameters (such as the electron
mass $m_0$ and charge $e_0$) are not measurable, we should not
care if they are finite or divergent, well-determined or
model-dependent. In a sense they do not exist since they are
certain abstract results of our imagination. What we have to care
about are only measurable quantities, i.e. `dressed' (perturbed)
parameters. We are able to express measurable energy shifts in
terms of the measurable electron mass $m$ and charge $e$ without
any divergencies and any needs for knowledge of physics at the
high momentum scale. This kind of expression of one measurable
quantity in terms of others means a QED calculation, a successful
QED calculation.

All these examples follow the idea of some equality between the
very existence of a quantity and the possibility for a measurement
of its value. This approach is a backbone of physics, its
operational philosophy.

It finds its realization in the approach of effective potentials,
which are used for various problems in particle physics. It may be
an effective phenomenological potential for pions, or an effective
quantum field theory produced on a way of going down to our energy
from the Planck scale or from the supersymmetry scale. The story
is that we believe that for various reasons the fundamental
physics is determined at certain much shorter distances and higher
energies and momenta than the ones we deal with in our
experiments. Dealing with low energies, we can see only a certain
effective theory. That is not a true fundamental theory but that
all what we have in an experimental sense. We have to be
successful, otherwise physics would have no sense until we reach
the fundamental scale of distances and energies. We trust that it
is enough to determine some parameters at our low energies and any
further calculations can be performed in their terms. In other
words, we expect that low-energy physics is complete (in a sense
that parameters determined at low energy are enough for the
low-energy calculations) and consistent. If that is not correct,
we should interpret that as existence of something unmeasurable
that affects our world in an unpredictable way.

\section{On Calculable Physical Constants\label{s:calc}}

\begin{flushright}
\begin{minipage}{9cm}
There was nothing so {\em very\/} remarkable in that; ... it
occurred to her that she ought to have wondered in this, but at
the time it all seemed quite natural.\\ \centerline{\em L.C.}
\end{minipage}
\end{flushright}

If we look at the list of the recommended values of the fundamental
constants \cite{codata}, it is unlikely to find there any constant
which can be calculated exactly {\em ab initio\/}. We can then
assume that there is no calculable constant which has a practical
sense. We have already mentioned in Sect.~\ref{s:in}, that considering
the hydrogen atom we either should deal with a calculable quantity
($R_\infty$), or with a measurable one ($\nu_H(1s-2s)$). If a
spectral property can be measured directly, it cannot be calculated
{\em ab initio\/} exactly. We may then conclude that there is no
constant which is both exactly calculable and directly measurable.

We note, however, that a reason for this conclusion comes in part
from psychology. Let us give an example of a similar situation. It
is well known that a theory of a point-like particle with a
non-zero anomalous magnetic moment is inconsistent. Meanwhile, we
believe, that the electron, being point-like, still possesses an
anomalous magnetic moment and its theory is consistent. The only
inconsistency here is in the terminology. We say ``the electron
has an anomalous magnetic moment", because it may be directly
measured and because it was first measured and next understood
theoretically. We say ``the electron is a point-like particle",
because its structure cannot be actually measured in any
straightforward way and because it was first calculated and next
certain previously measured effects (such as the Lamb shift in
hydrogen atom and helium ion) were understood as consequences of
the internal structure of the electron. From the theoretical point
of view the same effects are responsible for the anomalous
magnetic moment of the electron and its internal structure and in
a sense we can speak either about a point-like electron with $g=2$
or about an electron which has both the anomalous moment and the
structure. But for historical and psychological reasons we have
chosen another way to express the situation.

A similar problem in terminology is for the calculability of the constants.
We know that, e.g., an internal angular momentum of Earth and Moon could
take an arbitrary values and their ratio is a kind of constant to
characterize our Earth--Moon system. If the internal angular
moment (spin) were measured for quantum objects (such as electrons
or atoms) before the appearance of quantum mechanics (still it is
hard to imagine how), we could be surprised that $S_e/S_p=1$.
Quantum mechanics would explain this constant. However, in reality, first,
a quantum theory of the angular momentum was created and next we
measured the spin (or rather interpreted some results as a
determination of the spin of an electron and a proton). In time of
quantum mechanics the identity $S_e/S_p=1$ is trivial, and now we
do not consider the ratio of the spins as a fundamental constant.

Another example is the famous Einstein's identity $E_0=mc^2$. This
equation appeared as a result of special relativity and was first
seen experimentally through a relativistic correction to the
kinetic energy. There was no way to measure it directly. Now, we
can measure the binding energy $E_B$ (of nuclei, such as the
deuteron, or even of atoms --- see, e.g., \cite{pritch}) and check
whether the mass of a bound system is the same as a sum of the
masses of its composites. And we indeed know and can now verify
experimentally that the mass is reduced by a value of $E_B/c^2$.
We study the mass and the binding energy as static properties and
do not need to perform any relativistic experiment to check
$E_0=mc^2$. Another possibility to reach $E_0=mc^2$ without any
relativistic experiments is to measure annihilation energy of
positronium. The energy is determined as the energy of two
gamma-quanta and the positronium mass is twice the electron mass
(with a correction due to the atomic binding energy). If that was
measured before the Einstein's relativity theory, we would write
it as $E_0=k_1\cdot mc^2$ and interpret the theory as a
calculation of $k_1=1$.

One more example is a comparison of properties of a particle and
its antiparticle (like, e.g., their charges, masses etc.). That is
a result of the CPT invariance which is a consequence of the
Lorentz invariance. In early time, even the very existence of
antiparticles was first proved theoretically and next discovered
experimentally. We may say that we are able to calculate the
electron-to-positron mass ratio and it should be unity.

As we see from the examples above, very often a question of
calculability of a constant is related to history and psychology:
we should first recognize a certain property as a constant of
Nature and next calculate it. Generally speaking, the most
fundamental constants such as the speed of light $c$ or the Planck
constant $h$ enter a great number of very different equations. If
any of these equations were discovered before the Einstein's
relativity and quantum mechanics, we should introduce a number of
constants $c_1, c_2...$ and $h_1, h_2...$ instead of two basic
constants $c$ and $h$. Reducing the numerous coefficients in
different equations to these two, we, in a sense, calculate these
constants stating $c_1=c_2=...=c$ and $h_1=h_2=...=h$ (as it
is discussed above for $k_1=1$). And actually, that is one of the
most likely situations in future for exactly calculable constants.

Perhaps, the most important example of a similar situation is related
to the `elementary electric charge'. We accept for practical
applications that the absolute value of the electron and proton charges
are the same. For example, the CODATA adjustment \cite{codata} does not
distinguish between the proton charge and the positron charge.
However, no theory, confirmed by the experiment, implies that.
Conservation of the electric charge only urges that
\begin{equation}
q_e+q_p = q_n + q_\nu\;.
\end{equation}
In other words a small disbalance of the electron and proton
charges is permitted if the neutron and/or the neutrino possesses a
small electric charge. In earlier time we believed that the
neutrino was massless and thus should be neutral since a massless
charged particle would cause certain problems in conventional QED.
Now we have learned that the neutrino has a non-zero mass, but it
is suspected that this is the so-called Majorana mass, which also
implies neutrality of the neutrino. However, we can say nothing
about the neutron (from a theoretical point of view). Meantime,
from experiment we know that \cite{PDG}
\begin{equation}\label{qeqp}
\frac{\vert q_e+q_p \vert}{q_p} \leq 1.0\times 10^{-21}\;.
\end{equation}
The various limitations on $\vert q_e+q_p \vert$ involve certain
assumptions and we have to be very careful with the results.
However, the orders of magnitudes are quite clear. Briefly
speaking, when we consider an interaction of two hydrogen atoms at
a long (in a macroscopic sense) distance, the gravitational
interaction is 37 orders of magnitude weaker than the
electromagnetic Coulomb interaction of two protons. That means
that, if ${\vert q_e+q_p\vert}/{q_p} \geq 10^{-18}$, the
electromagnetic H-H interaction would dominate over the gravity.
We know that the interaction of `neutral' bulks of substance
apparently is the Newtonian's gravitation. If we suggest for
simplicity that $q_\nu=0$ (what is most probably true), then a
small value of $q_e+q_p = q_n$ would effectively produce an
$1/r$-force coupled to the baryon charge of the bulk matter. We
know (from various tests of the equivalence principle) that this
force (if any) is substantially weaker than Newtonian's gravity.
And that sets a limit on the residual electric charge of the
`neutral' hydrogen atom (and of the neutron) at the level of few
orders of magnitude below $10^{-18}q_p$. We note that the
limitation in (\ref{qeqp}) is only approximately three orders of
magnitude stronger than the limit of $10^{-18}q_p$ related to the
dominance of the gravity in the interaction of the `neutral'
particles. That is because of two reasons: first, the measurements
are related to the coupling constant, which is proportional to
$(q_e+q_p)^2$, and secondly, the mass itself is approximately
proportional to the baryon charge. Only small corrections, due to
a difference $(m_p+m_e)-m_n$ and a nuclear binding energy, violate
the equation $M_{\rm atom}=A m_n$. That considerably weakens use
of the equivalence principle.

If we believe in a certain unification theory (such as, e.g.,
SO(10)), we can {\em derive\/}
\begin{equation}
q_e+q_p =0\;.
\end{equation}
So considering different unification theories, we are approaching
a calculation of $q_e+q_p$, but it is likely that, once we
succeed, we will (for psychological reasons) say again ``that is
not a calculation since it is a trivial consequence of the
unification theory."

Let us return to the Rydberg constant. Have we calculated anything
real? Or did we just give a special name to a certain
experimentally meaningless combination of $e$, $c$, $h$ and $m_e$?
To answer this question we need to consider one more approach for
the {\em ab initio\/} calculation of properties in terms of the
fundamental constants: an approximate calculation. Such a
calculation is quite important for applications since we apply a
perturbative approach to numerous problems. Historically, the
Rydberg constant was introduced to describe certain hydrogen
energy levels (the Balmer series) and this constant was later
calculated. However, with a substantial increase of accuracy of
theory and experiment we arrived at a point when a choice had to
be made: to deal with a measured quantity or to introduce a
special value which would be used in perturbative calculations.
So, at the present time, the constant itself is not a real
property of any atom, and we can say that we gave a special name
to a specific combination of the more fundamental values. However,
we can say that there is a constant which describes (without any
corrections made) the hydrogen and deuterium energy levels with an
uncertainty below a part in $10^3$ and this constant has been
calculated. We may introduce certain corrections and bridge the
Rydberg constant and the hydrogen energy levels with much higher
accuracy. But the significance of the application of Rydberg
constants is first of all that this constant approximately
describes a number of transitions and that is an important success
of the {\em ab initio\/} calculations.

A significance of this constant is also that it describes the order of
magnitude for any gross-structure transition in any neutral atom and
molecule (see Sect.~\ref{ss:scal}). A constant characterizing an
effect in general is also an important and non-trivial result.
That is one more facet of the calculability of natural constants.
A famous example of such a calculation of order of magnitude is a
consideration by Schr\"odinger of the size of
atoms and the life cells \cite{life}. He tried to answer a
question, why the atoms are so small. Indeed, that is rather a
question why we are so big. Sch\"odinger considered some reasons
for that. This consideration has also been important to understand
the numerical values of the atomic constants, since the practical
units and, in particular, the SI units are defined in such a way that
anything related to a human being should be in a sense of order of unity.

Other examples: a prediction of the order of magnitude of the electrical
voltage which arose from molecular and atomic phenomena: it is a few
volts --- that takes its origin from early studies of similar phenomena
and from the mentioned above fact that all molecular and atomic energy levels
in the neutral atoms have the same order of magnitude (related to the
Rydberg energy which is approximately 13 eV). Actually, because of
that, the volt is only a natural unit if we mean its order of
magnitude.

{\em Ab initio\/} calculations in the leading order, or even a
rough approximation, may be also useful when one looks for a possible time
variation of the fundamental constants. In such a case it is necessary to
be able to perform a calculation of the dependence of energy levels on
the fundamental constants rather than the energy levels themselves.
We discuss this issue in Sect.~\ref{s:search}.

\section{Natural Units\label{s:unit}}

\begin{flushright}
\begin{minipage}{8.5cm}
And it certainly {\em did\/} seem a little provoking (`almost as
if it happened on purpose,' she thought).\\ \centerline{\em L.C.}
\end{minipage}
\end{flushright}

How many units and standards do we need? Theoretically, we need
only the base units of SI, and even not all of them. In particular, we can
reproduce the metre through the second and the ampere through the
kilogram, the metre and the second. However, practically, we need
a lot. A measurement is a comparison, and the most fortunate case
is a comparison of a quantity under question with a `probe
quantity' of the same kind. However, when we do different
measurements of `the same' kind of quantities such as, e.g.,
the distances, we notice a big range of their values. Astronomical and
atomic distances are related to not quite `the same' kind. And
indeed, for obvious practical reasons, we measure them quite
differently and like to apply different `probe quantities', i.e.,
different units.

Indeed, these units cannot be independent and we need to properly
calibrate them. However, the calibration is not always important,
because quite seldom we are really interested in a comparison of,
e.g., the mass of a hydrogen atom and the mass of the Sun. And
because of that we can leave their units, which must be related
in a formal sense, to be practically unrelated.

Still, for a number of measurements a proper calibration is
needed. In the case of classical phenomena we have to perform this
calibration regularly, to take care that the unit is unchanged
during the experiment etc. We should also take care that the units in
different laboratories are properly compared. However, quantum
physics opens another option. It offers quantum natural units
which are stable and universal.

Actually every dimensional fundamental constant is a kind of a
natural unit \cite{okun}, and a substantial part of dimensionless
constants (and certain dimensional constants such as the Boltzmann
constant $k$) can be considered as conversion factors\footnote{We
do not mean that these constants are just the conversion factors
and nothing else.} (see \cite{UFN}). So, there is a big variety of
various natural units and even natural systems of
units\footnote{We note that what is customarily referred to as
``systems of units'' is in fact a system of units and quantities.
For instance, the rationalized and irrationalized CGSG systems
deal with the same electrical units, but with differently defined
quantities. Natural systems of units sometimes assume
normalization of certain quantities different from the SI.}. Two
kinds of natural systems have been used.
\begin{itemize}
\item First, a system can have natural units for any dimensions,
such as the atomic units and the Planck units. Indeed, some systems are
incomplete because they do not care about all phenomena. But every
unit, which is really needed for a description of a certain class
of phenomena, may be or may be not related to the fundamental
constants. In the first kind of the natural units all necessary
units are related.
\item Secondly, a system can apply natural constants together with
other units. In such a case, the natural parameters or the fundamental
constants rather set a certain constraint on units, such as in the
case of systems in which $\hbar=c=1$ (relativistic units) or
$\epsilon_0=1$ (e.g., Gaussian units).
\end{itemize}

Are natural units (or a natural system of units) a good choice? In
their `complete form' they are as good as any other units.
However, in physics we widely use various `jargons' \cite{okun}.
We can indeed measure similar (but not the same) quantities in the
same units, such as a measurement both of the time intervals and
the distances in seconds (or both in metres). However, we know
these are very different quantities which are measured
differently. That means that saying $c=1$ we use a jargon, and in
reality, we mean something like $c$ is equal to one light year per
year or so. A jargon, as a special kind of language, differs from
a normal language being designed for a special use only. In this
special use (for a special kind of phenomena) it offers a more
short and clear description. Meantime, often the very use of the
`words' differs from the normal use and the jargon sentences are
`wrong' or meaningless literary. The same in physics. We like to
measure frequency, energy, momentum and mass in different units in
a general case. They are closely related in the case of
relativistic quantum physics. However, in the general case they
correspond to very different properties and assume different
experimental techniques to deal with them. They also suggest
different modifications for applications to continuous media
(which is rather unimportant for fundamental physics, but
significant for experiment). The practice of the jargons often
deal with numerous hidden substitutions for quantities
(confusingly keeping their names) such as
\begin{eqnarray}
t &\to& x_0 = ct\nonumber\\
m &\to& E_0 = mc^2\;.
\end{eqnarray}
Such hidden substitutions, which are equivalent to a use of the
same units for different quantities (as the energy units for the
mass), would be misleading and not very helpful in a general case
because of destroying advantages of the dimensional analysis
method. We like to distinguish the distinguishable quantities.
However, in quantum relativistic physics, where constants $c$ and
$h$ may appear in any equations, we can hardly use the dimensional
analysis and it is worth to present all of these quantities in,
e.g., energy units without losing even a bit of information.

Atomic units
\begin{eqnarray}
m_{\rm a.u.}&=& m_e =
9.109\,...\times10^{-31}\;\mbox{\rm kg}\nonumber\\
e_{\rm a.u.}&=& e = 1.602\,...\times10^{-19} \;\mbox{\rm C} \;,\nonumber\\
l_{\rm a.u.}&=& a_0=\frac{\hbar}{\alpha m_e c} =
5.291\,...\times10^{-11} \;\mbox{\rm m}
\;,\nonumber\\
E_{\rm a.u.} &=& E_h = \alpha^2 m_ec^2 = 4.359\,...\times10^{-18}
\;\mbox{\rm J}
\;,\nonumber\\
t_{\rm
a.u.}&=&\frac{\hbar}{E_h}=2.418\,...\times10^{-17}\;\mbox{\rm
s}\;,
\end{eqnarray}
which present a very natural, physical and logical coherent system
of units, are very well adjusted to atomic and molecular phenomena
and most of quantities there are of the order of unity. However, those
units indeed are not convenient for other phenomena. These units
present a case of the `theoretically' natural units. We choose them to
simplify theory (see, e.g., Sect.~\ref{ss:scal}). The other kind of
natural units are `practical' natural units. Choosing them, we do
not care too much about their fundamentality. Our concern is our
ability to apply them. Examples of the `practically' fundamental units
are the caesium HFS interval, the carbon atomic mass, the Bohr and
nuclear magnetons, the masses and the magnetic moments of an
electron and a proton, constants of von Klitzing and Josephson. We
partly consider a question of the `practically' fundamental units in
the next section (Sect.~\ref{s:mise}).

A choice on units, we do in physics, is quite simple. We use various
`theoretically' natural units when we do calculations. Some of them
are very helpful also for education. The more complicated are the
calculation, the more useful are the related natural units.
However, once we refer to a quantity to be measured, we switch to
`general' (SI) or natural `practical' units.

\section{Definitions and {\em Mise en Pratique} for the SI Units:\\ A back door for natural units\label{s:mise}}

\begin{flushright}
\begin{minipage}{9cm}
`When {\em I\/} use a word,' Humpty Dumpty said in rather a
scornful tone, `it means just what I choose it to mean
--- neither more nor less.'\\ \centerline{\em L.C.}
\end{minipage}
\end{flushright}

Currently, practical recommendations issued by CIPM for the most
important units \cite{SI} such as the metre \cite{meter}, the ohm
\cite{Ohm901,Ohm902} and the volt \cite{Volt90} are based on
certain natural units.

Why do we need such recommendations? A problem is that the
original SI definitions \cite{SI} cannot be used in an easy way.
As we mentioned, the idea of the units comes once from the fact
that a measurement is a comparison and to compare the results of
different measurements we need to go through a chain of
comparisons. The introduction of the units means that an essential
(and the `universal') part of the comparisons is separated from
the rest and recognized in a very specific way. It is a
responsibility of metrological institutions around the world to
take care about the standards and the units. The output of this
work should be a certain set of quantities, convenient for a
further use. Unfortunately, the SI definition of certain units is
not suited for that. The practical recommendations are designed to
cover the gap between the rigorous SI definitions and practical
accessability by a relatively broad range of users. However, the
recommendations are not a part of the SI in a sense: they aim to
arrange additional conventional units and to simplify a
measurement in the SI units as long as the users agree with a
reduced accuracy.

Let us give an example of such a recommendation. As we mention in
Sect.~\ref{ss:macro}, certain macroscopic quantum effects (the
quantum Hall effect and the Josephson effect) may be very helpful
to establish natural units of the resistance and the electric
potential. For that one has to know values of the fundamental
constants $R_K$ (the von Klitzing constant) and $K_J$ (the
Josephson constant) in the units of the SI. That is not a simple
issue and three basic strategies may be applied to take advantage
of these effects.

\underline{Scenario \# 1} suggests, that we fix values of these
two constants. Since that is not possible for the units of the SI,
that means an introduction of certain conventional units, in which
\begin{eqnarray}\label{eq:90}
R_K &=& 25\,812.807\;\Omega_{90}~~~~{\rm (exactly)}~~~~~~\protect\cite{Ohm901} \;, \nonumber\\
K_J &=& 483\,597.9\;\mbox{\rm GHz/V}_{90}~~~~{\rm (exactly)}~~~~~~
\protect\cite{Volt90} \;.
\end{eqnarray}
They are not the SI units and that means that if we check the
Ampere law with magnetic constant
\begin{equation}
\mu_0=4\pi\times10^{-7}\; \mbox{\rm N/A} ^2
\end{equation}
we should fail. We would also arrive at a discrepancy in the
energy measurements since
\begin{equation}
\mbox{\rm V}_{90}^2/\Omega_{90} \neq \mbox{\rm kg}\cdot\mbox{\rm
m}^2/\mbox{\rm s} \;.
\end{equation}
But for most of applications these mismatches with the SI system are not
important and the units volt-90 and ohm-90 are sufficient for
most of measurements, however, not for all.

If we really need to deal with the SI units, we should do
something else. \underline{Scenario \#2} suggests, that we use the
same units based on these two quantum effects, however, we
determine two necessary parameters $R_K$ and $K_J$ from additional
experiments. For trade and legal applications one may use the
recommendations \cite{Ohm902,Volt90}, which suggest an uncertainty
of the numerical values of (\ref{eq:90}) as related to the SI
units
\begin{eqnarray}\label{eq:90si}
R_K &=& 25\,812.807\,0(26)\;\Omega~~~~[1\times10^{-7}]~~~~~~\protect\cite{Ohm902} \;, \nonumber\\
K_J &=& 483\,597.9(2)\;\mbox{\rm GHz/V}~~~~[4\times10^{-7}]~~~~~~
\protect\cite{Volt90} \;.
\end{eqnarray}

Two scenarios above are based on the recommendations. The
recommendations, however, as well as all the legal metrology, are
designed not for a scientific use. What is important for physics
is not a subject of any legal agreement. For non-precision
absolute measurements, or for relative measurements, one can use
the CIPM \cite{Ohm901,Ohm902,Volt90} or CODATA \cite{codata}
recommendations just for convenience. Meanwhile, for precision
scientific applications, we should avoid any particular values of
$R_K$ and $K_J$ in the SI units at all. Instead, the results
should be presented as related to more complicated values, which
contain factors $(R_K)^n(K_J)^m$, taking into account that we have
measured a certain quantity in the units related to the quantum
natural units, determined by $R_K$ and $K_J$. One can see such an
approach in the CODATA adjustment of the fundamental constants
\cite{codata}, which dealt with the most accurate measurements of
the fundamental constants.

\section{Fundamental Constants and Geometry\label{s:geo}}

\begin{flushright}
\begin{minipage}{9cm}
... The Multiplication Table doesn't signify: let's try Geography.
\centerline{\em L.C.}
\end{minipage}
\end{flushright}

Speaking about the constants of Nature, we cannot avoid a question if
number $\pi$ is one of them. Our answer is: in a sense, it is. To
present our point of view we address to geometry. We remember from
high school, that geometry is based on twelve axioms, the
statements which are above any proofs. However, in physics we
should prove (experimentally) everything. We know that the general
relativity states that space-time is flat if there are no
gravity sources around. However, a correct statement is `locally
flat'. Globally, the universe may have a geometry which does not
allow the flat geometry `universewide' (like, e.g., a surface of
sphere). We have to check if the actual geometry is flat and the
present point of view is: it is close to being flat within a certain
uncertainty. That does not help much, because according to the
inflatory model \cite{Linde,Okun} the related parameters should be
very close to the values for the flat case. Topologically, the universe may be
either open or closed (or even of a more complicated topological
structure than just a closed 4D space). The closure or openness is
closely related to the density of energy which is known at a
few-percent level (see Sect.~\ref{s:cosm}). It is consistent with
the flat value. Now, we may in principle check whether the sum of
angles of a triangle studied in our space (after doing corrections
to remove gravity effects) is equal to $\pi$. This statement may
be verified and thus $\pi$ is a fundamental constant of our space.
However, it is a fundamental constant not in the same way as, e.g.,
$\alpha$. The fine structure constant is (at least now) a
constant without any relation to theory (of its origin). We
measure $\alpha$ and use its value to develop a QED phenomenology.
A value of $\alpha$ is in a sense not critical. If it were
discovered that $\alpha=1/136$ or $1/138$, that would certainly
change numerically the theoretical predictions, but would not
change our general view on fundamental physics. For the number
$\pi$ or the number of the space dimensions ($d=3$) we already have a
theory sensitive to their values. We can indeed say that $\pi$ or
$3$ are just mathematical numbers, imbedded into a flat $3D$
geometry. However, a physical statement is that this geometry is
ours. Actually, to measure the angles geometrically or
trigonometrically means to accept a part of geometrical ideas. In
such a sense any test of the sum of angles to be $\pi$ is rather a
test of the validity of the $3D$ Euclidean flat geometry in the application to our world.

The relation between geometry and predictability of the
fundamental constants has a more broad context. In a sense any
symmetry is related to a kind of geometry.

Trying to build relativistic gravity theory, we strengthen
significance of the equivalence principle which says that a ratio
of the inertial and gravitation masses is a universal parameter
(which we set to unity).

A wish to build a linear equation for a relativistic particle led
to a prediction of positrons with `calculable' properties:
\begin{eqnarray}
q_{\overline{e}} &=& -q_e \;,\nonumber\\
m_{\overline{e}}  &=& m_e \;, \nonumber\\
I_{\overline{e}}  &=& I_e \;,\nonumber\\
\mu_{\overline{e}}  &=& -\mu_e \;.
\end{eqnarray}
The positron actually has been the first ever predicted particle.
The Dirac equation (as well as the other linear equations)
suggests that in the leading approximation the $g$ factor of a
point-like (i.e. structureless) particle is equal to two. However,
this value is perturbed. For the free leptons (electrons and muon)
the non-zero anomalous contribution is small and can be
perturbatively calculated up to certain accuracy and also
accurately measured. The present combined results for the
anomalous magnetic moments are \cite{codata}
\begin{eqnarray}
\left(\frac{g-2}{2}\right)_e &=& a_e = 1.159\,652\,185\,9(38)\times10^{-3}~~~~[3.2\times10^{-9}]\;,\nonumber\\
\left(\frac{g-2}{2}\right)_\mu &=& a_\mu =
1.165\,919\,81(62)\times10^{-3}~~~~[5.3\times10^{-7}]\;.
\end{eqnarray}

When the unification theory SU(5) was suggested, one of its
successes was an explanation of the Weinberg angle $\Theta_W$. The
value was predicted for a certain high energy scale (related to
$E_{\rm un}\sim 10^{14}\;$Gev). A measurable value is related to a
much lower energy scale and thus should be renormalized. An
accurate theory of its perturbation by the radiative corrections
together with accurate experimental data should provide us with a
constraint on the unification theory. The SU(5) theory happens to
be incorrect since it disagrees with a number of observed effects
(such as a `too long' proton lifetime $\tau_p$, presence of the
neutrino oscillations etc.). It is believed, nevertheless, that a
similar unification scheme will eventually explain the neutrality
of the hydrogen atom and the value $\Theta_W$.

It is quite likely that most, if not all, calculable constants may
be predicted only via symmetrical and thus geometrical ideas.

The symmetries and conservation laws are very strongly related to
the fundamental constants and their constancy. Due to that we would like
to note that additionally to direct violations of symmetries
quantum physics offers additional ways to violate the symmetries in
a `smooth' matter: via the quantum anomaly or the spontaneous breaking.
\begin{itemize}
\item[{\em anomaly\/}] Quantum field theory suggests that we
should substitute the wave function for a field operator
$\Psi(x)$, which is quite singular and a quantity
$J[\Gamma]=\overline{\Psi}(x)\Gamma\Psi(x)$ (where for fermions
$\Gamma$ is an arbitrary combination of the Dirac gamma-matrices)
is ill-defined. The quantity $J$ is significant because any
electron or quark current is of such a form. It was discovered
that there may be a special kind of quantum violation of symmetry
--- the anomaly. It is realized in such a way that:
\begin{itemize}
\item[(i)] there are two currents $J^1_\mu$ and $J^2_\mu$, which
are conserved within classical physics ($\partial J^a_\mu/\partial
x_\mu=0$), and thus two symmetries are presented at the classical level;
\item[(ii)] the
currents are singular and thus ill-defined at the quantum-field level;
\item[(iii)] there is no
regularization which supports the conservation of both currents and
both symmetries.
\end{itemize}
As a result, one of the symmetries (e.g., related to
$J^2_\mu$) is to be violated and the non-conservation term
$\partial J_\mu/\partial x_\mu\neq0$ is proportional to the Planck
constant $h$ (see \cite{Itzykson} for more detail). A well-known
example is the Adler anomaly for the axial current which plays
an important role in properties of the $\pi^0$-meson.
\item[{\em spontaneous\/}] The spontaneous breakdown of symmetry
is another example how a classical symmetry can be broken in
quantum field theory. Let us suggest that the interactions
(potentials) are invariant in respect to a certain symmetry. There
is no symmetrical state with the minimal energy, but instead there
is a family of non-symmetrical minimum-energy states. It is
similar to, e.g., the magnetization of a bulk of iron. The theory
is isotropic, however, a minimum of energy is related to the case
with a certain non-zero value of the macroscopic magnetic moment.
Any direction of the moment is related to a minimum of the energy
(`vacuum'), however, only one can take place at any particular
case. Indeed, there are domains with different directions, but if
our observable universe is inside such a domain, we would not see
the other domains. We note that it is different from the simplest
problems in quantum mechanics. Quantum mechanics can also deal
with such a potential, however, because of the overlap of the
vacuum states there is a `fine structure' and the actual minimum
of the energy is related to a certain superposition of these
states (e.g., their symmetric sum). The particular asymmetric
vacuum state is to be presented as a sum of the superposition of
compound vacuum states and its evolution via certain oscillations
will lead most probably to the lowest state of this fine
structure. However, with an increase of the phase volume (with
increase of the number of degrees of freedom) the probability of
the tunnel transitions between the vacuum states goes down very
fast. The evolution time becomes so long that we can see no
evolution et all. E.g., we cannot detect any oscillation between
left-hand and right-hand organic molecules. In the case of the
quantum field the characteristic evolution time is infinite
because of the huge volume of the universe (see
\cite{okunbook,Okun,Itzykson,Linde} for more detail).
\end{itemize}

There is also a specific kind of phenomena, which may lead to an
`observational' violation of such symmetries as, e.g., the Lorentz
invariance. They are related to the fact that it is unlikely to
observe any symmetry directly, but we study certain consequences
of the symmetry, and, if we do not know a complete theory, we can be misled.

Let us explain that with an example of nonrelativistic quantum and
classical mechanics. They have mainly the same symmetries and
conservation laws (conservation of momentum, angular momentum and
energy). However, from a classical point of view, conservation
means that we can measure, e.g., all three components of the
angular momentum in two separate moments of time and the result
must the same. Classically, we also expect that we can do two
`fast' separate measurements of the energy, as precise as we like,
and that allows to check whether the energy is conserved exactly
and in any particular phenomena or the conservation takes place
approximately and/or on average. And `fast' means that the
measurement time may be as short as we like. That is by far not
the same in the quantum case. In both cases, doing `classical'
experiments for the conservation of the `whole' angular momentum
and the `exact' conservation of energy, one will fail to confirm
the conservation laws once we arrive at the level of accuracy
where quantum effects enter into the game. The symmetries and
conservations still take place, but their observed consequences
differ from the naive classical expectations.

We know, that at the Planck scale, the geometry of the space-time
quite likely differs from what we see around us\footnote{I have
heard that numerously and in particular a statement about a
``non-commutative space'' and this is one more example of a
physical jargon. We, e.g., clearly understand that quantum
mechanics does not change the phase space. In the one-dimensional
case, the plain $\{x-p\}$ is just the same as in the classical
case. However, classical states are rather point-like. We
sometimes assign them a finite volume because of the experimental
uncertainty in our data or because of their statistical treatment
which is in classical physics also a result of an uncertainty in
our description. Meanwhile, the volume presents only a spot where
the point may be, but any state still is point-like and the
uncertainty is, in principle, avoidable in classical physics. The
Heizenberg inequality implies that a quantum state has a minimal
finite volume determined by the Planck constant $\hbar$. The same
for the $3D$ angular momentum. The only point-like quantum state
in the $3D$ angular-momentum space corresponds to the case of zero
angular momentum. But nevertheless
--- the space is the same, the allowed states are different.}. We
do not know what it really is. Indeed, certain symmetries can be
broken there. However, some most `sacred' symmetries might be
realized in such a way that their consequences alternate from our
expectations and, thence, the experimental results might `observe'
certain violations of these symmetries and conservations.

\section{Constancy of Fundamental Constants\label{s:const}}

\begin{flushright}
\begin{minipage}{9cm}
... They began running when they liked, and left off when they
liked.\\ \centerline{\em L.C.}
\end{minipage}
\end{flushright}

The fundamental constants, most of them, appear in physics with
quantum mechanics. The Newton's constant $G$ came earlier, but
only considering the Planck-scale effects we can imagine how
fundamental it is. They were called `constants' and it was
believed that they should be such by default. To vary them, one
should rather expect an exceptional reason. That was the
situation, when Dirac and later Gamov suggested that the
`constants' may not be constant.

However, the truth is that there is no strong reason why the
`constants' of Nature are constant. We know that the ratio of the
electron and proton spins is unity and cannot vary. If it were
possible to switch off the QED corrections, we should expect that
the $g$ factor of an electron is a trivial constant equal to two.
Thus, there may be only one theoretical reason for their constancy
--- that would be an explanation of their origin. For the most
important constants we have none. The constancy of the constants
is merely an experimental fact and an {\em a priori\/} trust in
the domination of symmetry in the nature of Nature. The former,
indeed, can never be final and we need to check that again and
again with a more broad range of phenomena and with a higher
accuracy. The latter is in a formal sense rather wrong: we
recognize the inflation as a basic element of modern cosmology.
And the inflation \cite{Linde,Okun} had urged the electron mass
and charge to vary in a very remote past. If we accept that the
constants were varying once, we should rather consider them as
changing quantities at a default situation, and need a reason for
them not to vary again. Or not to vary fast. An once non-constant
is forever not a `trusted' constant.

We recognize the existence of the dark matter which may interact
`very' weakly with our matter. We do not know what the dark matter
is and how weak may be this `very' weak interaction. Due to a
number of such unclear phenomena, we need to distinguish between
\begin{itemize}
\item effects such as a violation of the local position invariance
(and in particular a violation of the local time invariance)
\item and a variation of the constants.
\end{itemize}
One may expect that a violation of the local invariance means
that results of measurements would depend on time and location
upon the measurement and that is the same as a time- and space-
dependence of the fundamental constants. However, these two
situations are not quite the same.

The results of an experiment may be affected by an environment. In
earlier times, an `environment' for a laboratory-scale experiment
was also laboratory-scaled. An exception was gravitational and
magnetic field of Earth. However, they were not significant: since
the former was nearly a constant (which did not depend on the
location at the level of then achievable accuracy) and from the latter
there may be a shield. Now, doing high-precision balance
experiments, one can clearly see effects of the motion of Sun and
Moon in this scale of experiments. Indeed, the existence of the
surf has been known for centuries. But the surf is a result of an
accumulation of these effects over a `big' detector, which is of
the Earth scale. Until the very recent time it was not possible to
see such effects with the `small' detectors.

Now, we are sensitive to the environment at a very large scale. We
know that we live in a changing universe (the environment item
number one), going through a bath of 2.7-K cosmic microwave
background and a similar background radiation of known (neutrino)
and, maybe, unknown massless particles (the environment item
number two) and dark matter and dark energy presented around (the
environment item number three) etc. We would never qualify any
effect of interactions with them as a real violation of Lorentz
symmetry, but we may want to qualify a variation of certain
natural parameters induced by them as a variation of the
constants. In principle, we can say that there was no variation of
truly fundamental constants during the inflation, but only
`environmental effects', caused by cooling of the Universe.
However, we prefer to say that the electron mass has changed.

As we mentioned, the Earth gravity field is nearly a constant and
the free fall acceleration $g$ was considered for a while as an
universal and fundamental constant. Now we know it is neither
constant nor universal and fundamental.

Kepler found that any planetary orbit satisfies a condition
\begin{equation}\label{e:kepl}
\frac{R^3}{T^2}=\mbox{\rm [Kepler's]~constant}
\end{equation}
with the same universal constant for any planet. We now know that the
Kepler's `universal' constant, which governs the motion of all planets, is
a specific constant related to our solar system only and nothing more.

These two examples shows how important it is to understand the
nature of the constants. We now have a great number of fundamental
parameters, origin of which is unclear: the Yukawa Higgs coupling
constants, the Cabibbo-Kobayashi-Maskawa matrix (CKM) parameters,
parameters of a lepton analog of CKM, cosmological constants etc.
These constants have been observed and studied. There are also a
number of important
constants which have not yet been detected, but strongly expected
as, e.g., the mass of the Higgs particle.

Albert Einstein believed that all the constants are, in principle,
calculable. That should be expected in a world, where the
equations determine everything. But that apparently is not ours.
We know, that some symmetries of our world have been spontaneously
broken. That happens, when the symmetric state is unstable, while
a family of non-symmetric states has the same minimal energy. The
vacuum falls to one of these minimum energy states. We know a
number of examples in classical physics. For example, we already
mentioned a bulk piece of iron with a non-zero value of the
residual magnetic moments. A massive piece used to consist of
domains --- numerous smaller pieces, in which there is a
non-vanishing macroscopic magnetic moment. Hamiltonian and all
equations which describe any domain are isotropic. However, the
state with zero macroscopic magnetic moment is unstable. Stable
states are those with a magnetic moment directed to somewhere.
Where? We cannot predict. It may be any direction and in fact the
directions in different domains are different. One may think that
a direction is not as important for observable quantities as the
magnitude (indeed until we do not look for a violation of the
isotropy --- which may be not important if we have in mind not our
space but a certain functional space like e.g. of the isotopic
spin). However, there is a simple example how to transform the
direction into a magnitude. It is enough to imagine a situation
when there are two independent values similar to the iron's
magnetic field, completely independent for the vacuum states, but
coupled together to the same matter field (i.e. to a certain
particle). In such a case the angle between them is related to a
scalar which can affect a value of the energy of the particle.

This example shows, that certain properties cannot be predicted
through the equations. We acknowledge a spontaneous breakdown of
symmetry in the Standard model of the electroweak interactions
\cite{okunbook,Okun,Itzykson}. We expect that our world has a
larger symmetry group than we actually observe. And a
non-observed part of the symmetry has been destroyed by one or few
spontaneous breakdowns. It may happen that certain `fundamental'
parameters of our world are a direct result of such breakdowns and
they could take, in principle, different values in another place or
another version of the evolution of the universe.

If that is the case, certain parameters are not predictable and
discussing them we approach a framework of the so-called anthropic
principle. There has been a number of various modifications of it,
including not only physical, but also philosophical ideas. We are
not very enthusiastic about these ideas. However, a `minimal'
physical part of the principle is the {\em selection principle\/}:
we observe only what we can observe and the very presence of our
species, as the observers, sets a certain constraint on the
observable properties. That is like the second Kepler's law: if we
would learn neither the Newtonian gravity theory, nor data about
planets outside of our solar system, we should consider the
Kepler's constant in Eq.~(\ref{e:kepl}) as a universal constant
--- the universal constant for all observable planets.

Once we allow variations of the constants of Nature, we remark that the
units are also vulnerable. From the first glance, we should prefer to
speak about dimensionless constants such as
\begin{equation}
\alpha=\frac{e^2}{4\pi \epsilon_0 \hbar c}\;.
\end{equation}
They are clearer to discuss phenomenologically and easier to
detect. Sometimes, it is even stated that we can only look for a
variation of dimensionless quantities.

Indeed, variations of the dimensional constants may be also
detected, but the experiments are much more complicated, because
they should directly address time- and space- gradients of such
constants \cite{eprints,ACFCi}. A well-known example is the famous
Michelson-Morley experiment, which checked whether the dimensional
constant (the speed of light) in the same in any directions.

Let us leave the general discussion on a search for the
variability of the constants at this point and first look how we
can describe their variations. If one tries to seriously consider
the varying constants, we have to introduce changes from the very
beginning. It is not enough just to accept the equations derived
under a conventional assumption of the constancy of the natural
constants and then allow them to vary slowly. We can easily arrive
at a contradiction. Let us consider a simple example: a situation
when, in a specific inertial frame, the Planck constant is slowly
changing globally with the time, so $h=h(t)$. Meantime, the laws
of physics are still isotropic and in particular $\partial
h/\partial {\bf x}_i = 0$. However, the angular
momentum\footnote{We consider here the classical angular momentum
${\bf L}$, which is dimensional, and the quantum angular momentum
${\bf l}$ which is dimensionless.} is quantized
\begin{equation}
L_z =  \hbar\cdot l_z\;.
\end
{equation}
We arrive at an obvious contradiction: the angular momentum should
be conserved (because of the isotropy), while $l_z$ is also not
changing (being an integer or semi-integer number) and, meantime,
their ratio, the Planck constants $\hbar$, is changing. This
inconsistency comes directly from the assumption that we can
accept the known equations and allow their constants to vary
slowly.

There is no straightforward way to deal with the variable constants.
First of all, when the constants are {\em constant\/}, we can
redefine operators via their effective renormalization as, e.g.,
\begin{equation}\label{e:eFmn}
F_{\mu\nu} \to \frac{1}{e}F_{\mu\nu}^\prime
\end{equation}
etc. If we like to introduce slowly varying constants into the
basic equations, we even do not know into which. Because of
the `renormalizations', such as in (\ref{e:eFmn}), we have not a
single set of
the basic equations but quite a broad family of the equations which are
equivalent as long as the `constants' are constant.

We used to describe most of quantities which are calculable with
help of such equations. However, in our opinion, the starting
point to adjust our basic phenomenology to the case of the
variable constants is the path integral (the functional integral
over field configurations). The conventional approach reads that
the path integral
\[
Z=\int{e^{-iS^\prime}}
\]
presents a matrix element of the evolution operator
\cite{Ramond,Itzykson}. To study a particular evolution we should
integrate over all available configurations (`trajectories') with
proper initial and final conditions. The action $S^\prime$ is
normalized to be dimensionless. This operator has a transparent
physical sense: we have to sum over all possible trajectories and
we also know that in most of cases the dominant trajectory
(trajectories) is related to the least action. The least-action
trajectory for quantum mechanics is the classical trajectory. When
we study quantum-field `trajectories' in the functional space the
least-action trajectory is related to the field equation such as
the Maxwell equations for the photon's field and the Dirac
equation for the electron's fields. Far from the minimizing
trajectory the phase (which is the action $S^\prime$) is changing
fast and the contributions cancel each other. Close to the
minimizing trajectory the phase is nearly unchanged and the
contributions are enhanced.

Now, we can generalize the action (e.g., for quantum
electrodynamics, which is in a narrowed sense, substantially, a
theory of electrons and photons --- see Sect.~\ref{S:simple}) to
the form
\begin{equation}\label{SQEDx}
S_{\rm QED}^\prime = \int{d^4x}\left\{ \xi_3(x)\overline\psi
\left[ g_{\mu\nu}\gamma^\mu \left(i\frac{\partial}{\partial x_\nu}
+\xi_4(x)A^\nu\right)
-\xi_5(x) \right] \psi -\frac{1}{4}\xi_6(x)g_{\mu\nu}
g_{\rho\lambda} F^{\mu\rho} F^{\nu\lambda} \right\} \;,
\end{equation}
where the metric tensor at a particular preferred
frame\footnote{There are two natural options for such a frame. The
first is related to the one in which the cosmic microwave
background (CMB) radiation is isotropic. The other corresponds to
the frame which is determined by the dark matter. This latter is
well defined locally, but not globally. Once we fix the frame we
can consider an analogy with electrodynamics in media, which in
the simplest case can be described by two dimensionless functions
$\epsilon_{\rm rel}(x)$ and $\mu_{\rm rel}(x)$.} is defined as
\[
  g_{\mu\nu}=
  \left(
  \begin{array}{cccc}
    \xi_1(x)&0&0&0\\
    0&-\xi_2(x)&0&0\\
    0&0&-\xi_2(x)&0\\
    0&0&0&-\xi_2(x)
  \end{array}
  \right)
\]
and the electromagnetic field tensor as
\[
F^{\mu\nu} = \frac{\partial A^\nu}{\partial x_\mu}-\frac{\partial
A^\mu}{\partial x_\nu}\;.
\]

The functions $\xi_n(x)$ obviously violate several important
symmetries (including the gauge invariance, the local position
invariance, the local Lorentz invariance) and allow time- and
space- variations of the dimensional fundamental `constants'.
Here, the functions $\xi_n(x)$ are dimensional, but we can indeed
introduce factors $h^{(0)}$, $c^{(0)}$, $e^{(0)}$ and $m^{(0)}$ in
a proper way and make the $\xi$ functions dimensionless. These
four factors are related to the Planck constant, the speed of
light, the elementary charge and the electron mass as measured in
a particular point $x^{(0)}$. More complicated models are also
possible with, e.g., a less trivial metric tensor or the
electromagnetic field tensor, or with appearance of small terms
which directly violate various symmetries.

When should we use Eq.~\ref{SQEDx} and when just put time- and
space- dependence inside conventional equations? It depends on a
problem. Basically, there are three kinds of related measurements:
\begin{itemize}
\item One may perform a series of `fast' measurements separated in
time by a `large' interval $T$  (for simplicity we speak here
about the time variations only). That is, in particular, a case of
atomic, molecular and nuclear spectroscopy. Additional terms with
derivatives of the $\xi(x)$ functions should be integrated over
the `short' time of the measurements $\tau$ and the value of
$\tau\partial \xi/\partial t$ may be neglected in comparison with
a difference $\xi(t+T)-\xi(t)$ over the `large' separation
($T\gg\tau$).
\item `Long' measurements can be performed, e.g., as a result of
continuous monitoring of the motion of planets etc. In this case
the effect of the integration of $\partial \xi/\partial x$ is
comparable to the effect of the adiabatic change of $\xi(x)$ in the
conventional equations.
\item And indeed, one can try to deal directly with derivatives
performing $\partial \xi/\partial x$-sensitive experiments.
It is quite probable
 that it is easier to do that for space- rather than for time-
gradients. For example, we can try to perform precision
measurements in space similar to the GPS measurements in the
atmosphere. In the case of space-gradient terms in the law of the
propagation of light we should `observe' a non-flat geometry after
interpreting the light propagation time intervals as the effective
distances.
\end{itemize}

\section{Search for Possible Time Variation of Fundamental
Constants\label{s:search}}

\begin{flushright}
\begin{minipage}{6.5cm}
-- ...`one {\em can't \/} believe impossible things.'\\ -- `I
daresay you haven't had much practice.'\\ \centerline{\em L.C.}
\end{minipage}
\end{flushright}

The easiest and most transparent kind of experiments to search for
a variation of the constants is indeed to measure the same
quantity twice. If these measurements are `fast' and have a long
separation, we can use the description with non-varying constants
and not care about possible additional terms and gradients. The
validity of this approach is obvious for atomic physics: we do a
series of short measurements, for which the gradients of constants
are negligible for atomic time and space scale and cannot affect
the result of measurements. A different situation is for a search
of a variation of the Newtonian constant $G$. Most of measurements
are related to continuous monitoring during a long period. Instead
of large series of short atomic measurements, the gravity searches
deal with a number of long measurements. In such a case,
contributions of the gradients should be important\footnote{It
should be understood, that varying $G$ and its gradients probably
is not enough. For instance, the equation for a photon in a
medium, presented in terms of vacuum fields, would involve not
$c(x)$ and its gradients, but instead $\epsilon_{\rm rel}(x)$ and
$\mu_{\rm rel}(x)$ and their gradients. Those functions have no
separate sense in vacuum.}.

Due to that we concentrate our attention on atomic measurements.
Still there are three kinds of them discussed in literature.
\begin{itemize}
\item Astrophysical comparisons deal with a relatively low
fractional accuracy, but take advantage of a very big time
separation (up to $10^{10}$ yr). This kind of observations is by far
not transparent and suffers from necessary statistical
evaluations in a situation when certain correlations may be
present.
\item Clocks based on different transitions have been developed
and comparing them one may hope to learn about relative variations
of their transition frequencies. However, the clock is just a
device, built for a purpose, and it is not necessary that its
properties are related to the atomic or molecular transition
frequency in an exact sense. There may be a number of drifts and
certain parameters of the clock, which determines drifts and which
are determined in turn by an artificial environment in a
non-controlled way. An example of such a clock is the hydrogen
maser. Its frequency drift is purely due to an environmental
problem (a so called wall-shift).
\item Still, the frequencies of certain clocks follow the atomic
transition frequencies. Such clocks are similar in this sense to
the primary caesium standard. The very basic principle of the
primary caesium clock is that it should reproduce the caesium HFS
transition frequency, which is related to the SI second. To deal
with this kind of the clocks is the same as to measure an atomic
or molecular frequency with high accuracy.
\end{itemize}
Below we consider in detail this kind of the `near primary'
frequency standards (and we note that maybe in future one of them
will give us a new SI second\footnote{Because of a certain
conservativeness of CIPM, which should necessarily take place, and
because of a variety of competitive optical candidates, we expect
in near future not a change of the definition of the SI second,
but, first of all, certain CIPM recommendations. At the first
stage, they could recommend values of certain optical and
microwave transitions which would be advised for a practical
realization of the second (compare, e.g., to the CIPM
recommendation on the metre \cite{meter}). At the second stage,
after the accuracy of a comparison of certain optical transitions
to each other will supersede the accuracy of the caesium
standards, a conventional second (e.g., the second-2015) could be
introduced.}).

Significance of these clock-based experiments is in a great
accuracy of precision frequency measurements. Presently and for a
while (maybe even forever) the frequency measurements are the most
accurate.

Here we consider certain detail of laboratory searches. Different
aspects of the possible variations of the fundamental constants
and their searches are discussed in \cite{ACFC}.

\subsection{Atomic clocks}

In this chapter we consider rather frequency standards than
clocks. In principle, a clock is a time standard. Indeed, the
frequency and time intervals are closely related, however, the
time measurement may be `absolute', i.e., related to the
conventional `beginning of time'. That involves two metrological
problems for keeping the time scale: the realization of the
time-interval unit and of the `zero point'. The specifics of time
keeping requires that `true' clocks operate continuously,
otherwise the information on the initial moment would be lost. A
real time standard is actually not a single device but a set of
various related standards. Still, with a peripheral part of the
clock operating around it, the very heart of any clock is a
certain frequency standard. Presently, that is either a caesium
standard or a standard calibrated against the caesium.

The best modern clocks pretend to deliver certain reference
frequencies with an uncertainty at the level of a part in
$10^{14}$ and even less. If we check a value of the linear Doppler
shift related to this level, the speed of an atom is to be
$3\;\mu$m/s. For a hydrogen atom a temperature of 1 K is related
to a speed of approximately 100 m/s, i.e., eight orders of
magnitude higher. Heavier atoms at this temperature are slower by
a factor of $\sqrt{A}$, where A is the atomic mass number. That
means that for an accurate clock we have to solve the problem of
the linear Doppler effect.

In different clocks the problem of the linear Doppler effects is
solved differently (see \cite{AMOP} for more detail). In the
clocks with neutral atoms, the atoms are cooled down to the level
much much lower than 1 K. Ions are trapped and that eliminates the
Doppler effect --- a localized trapped particle cannot have a
non-zero momentum. One more approach is to study two-photon
transitions, which are not sensitive to the linear Doppler effect.

If the constants are changing, not only theory should be
reconsidered, but also experiment. First of all, we need to
acknowledge that if certain natural constants are changing, our
units and, in particular, the SI second are changing as well. For
this reason, the most simple and hopeful way is to look for
variation of dimensionless quantities. However, since any
measurement is a comparison, we can also deal with dimensional
quantities, properly specifying the units. The interpretation of
the variation of the numerical values of the constants differs
drastically from the interpretation of a search for their direct
variation. In particular, we will speak about a variation of the
numerical value of the Rydberg constant, which is closely related
to properties of the caesium atom. In fact, it is equal to
\begin{eqnarray}
\{R_\infty\} &=& \frac{2c\cdot R_\infty}{\nu_{\rm
HFS}({}^{133}{\rm Cs}) } \times \frac{\nu_{\rm HFS}({}^{133}{\rm
Cs})}{2c}
\nonumber\\
&=& \frac{1}{\bigl(\nu_{\rm HFS}({}^{133}{\rm Cs})\bigr)_{\rm
a.u.}}\times \frac{9\,192\,631\,770}{2\times
299\,792\,458}~~~~{\rm (exactly)}
\end{eqnarray}
where $\bigl(\nu_{\rm HFS}({}^{133}{\rm Cs})\bigr)_{\rm a.u.}$ is
the caesium HFS interval in the atomic units and the exactly known
number
\[
\frac{9\,192\,631\,770}{2\times 299\,792\,458} =
15.331\,659\,494\,249\,183\,8\dots
\]
is an artifact of the SI system.

\subsection{Scaling of different transitions in terms of the fundamental constants\label{ss:scal}}

With help of the accurate clocks we can compare the frequencies of
different transitions. What can we learn from them?

First of all, let us look at expressions for different atomic
transitions in the simplest case, i.e. for the hydrogen atom,
\begin{eqnarray}\label{eq:scal}
f(2p-1s)&\simeq&\frac{3}{4}\cdot cR_\infty\;,\nonumber\\
f(2p_{3/2}-2p_{1/2})&\simeq&\frac{1}{16}\cdot\alpha^2\cdot
cR_\infty\;,\nonumber\\
f_{\rm HFS}(1s) &\simeq&\frac{4}{3}\cdot\alpha^2\cdot
\frac{\mu_p}{\mu_{\rm B}}
 \cdot cR_\infty\;.
\end{eqnarray}
Indeed, there are various corrections and, in particular, the
relativistic and the finite-nuclear-mass corrections but they are
small.

The first interval is related to the gross structure, the second
is for the fine structure and the last is for the hyperfine
splitting. So, we note that if we would measure them, we can learn
about variations of $cR_\infty$, $\alpha$ and $\mu_p/\mu_B$. To be
more precise, when one measures a frequency, the result may be
either absolute or relative. The latter case, when two ratios
are measured
for three transitions, will tell us nothing about the
Rydberg constant. Measuring the intervals absolutely, i.e. in
certain units, we can consider a variation of a value of the
Rydberg frequency $cR_\infty$ in these units. In the previous
subsection we explained about the physical meaning of a value of the
Rydberg constant in the SI units.

What happens if we consider a more complicated atom? First,
let us re-write the equations above in atomic units
\begin{eqnarray}\label{eq:scalau}
f(2p\to 1s)\Bigl\vert_{\rm a.u.}&\simeq&\frac{3}{8}\;,\nonumber\\
f(2p_{3/2}-2p_{1/2})\Bigl\vert_{\rm
a.u.}&\simeq&\frac{1}{32}\cdot\alpha^2\;,\nonumber\\
f_{\rm HFS}(1s)\Bigl\vert_{\rm a.u.}
&\simeq&\frac{2}{3}\cdot\alpha^2\cdot \frac{\mu_p}{\mu_{\rm B}}\;.
\end{eqnarray}
The gross structure is of order of unity. The fine structure is a
relativistic effect, proportional to the factor of $(v/c)^2$ and thus to
$\alpha^2$. The hyperfine structure is also a relativistic
effect, but it suppressed by a small value of the nuclear
magnetic moment in atomic units.

If we have a more complicated atom, nothing will change except for
the numerical coefficients. There is no additional small or big parameter
when calculating the gross structure and it still should be of order of
unity. The electron speed is always proportional to $\alpha c$.
It may also be a value of the nuclear charge $Z$ involved, but it
does not change with time. So, we conclude that the scaling
behavior of the atomic transitions with changes of the constants is the same as in
hydrogen (see Table~\ref{t:scal-a}). The importance of these
scalings for a search of the variation of the constants was first pointed
out in \cite{savedoff-99} and discussed there for
astrophysical searches.

\begin{table}
\begin{center}
\begin{tabular}{l|r}
\hline
Transition & Energy scaling \\
\hline
Gross structure & $ cR_\infty $ \\
Fine structure & $\alpha^2 cR_\infty $ \\
Hyperfine structure & $\alpha^2(\mu/\mu_B) cR_\infty$ \\
\hline
\end{tabular}
\end{center}
\caption{Scaling behavior of the atomic energy intervals as functions
of the fundamental constants. $\mu$ stands for the nuclear
magnetic moment. \label{t:scal-a}}
\end{table}

Molecular spectra are more complicated than atomic. The biggest
energy intervals are related to the electron transitions and they
are completely similar to the atomic gross structure. Two other kinds
of the intervals (vibrational and rotational) are due to the nuclear
motion.

Let us consider a diatomic molecule. In the so-called
Born-Oppenheimer approximation (see, e.g., \cite{molbook}) we can
consider the energy of the electronic states as a solution of a
problem of the electrons in the field of two Coulomb centers with
the infinite masses separated by the distance $R$. The result
depends on this distance ($E(R)$) and the next step is to find a
value of the distance $R_0$ which minimizes the energy. Let us now
take into account nuclear motion. In the leading approximation the
Hamiltonian is of the form
\begin{eqnarray}
H &=& \frac{{\bf P}^2}{2M} + E(R) \nonumber\\
&\simeq& \frac{{\bf P}^2}{2M} -\frac{k}{2}(R_0-R)^2+ E(R_0)\;,
\end{eqnarray}
where we note that $R_0$ is about unity in atomic units, the
binding energy $E(R)$ is also about unity and thus $k\sim
E(R)/R^2_0$ is about unity as well. All of them do not depend on
the fundamental constants (in the atomic units). $M$ is the
nuclear reduced mass. The equation is for a harmonic oscillator
(in the leading approximation) and we know all parameters (at
least their dependence on the constants in the atomic units). We
find that the vibrational quantum of energy is $\sqrt{1/M}$ in the
atomic units or $(m_e/M)^{1/2}cR_\infty$ in the SI units. Here $M$
is a characteristic nuclear mass, but for the most of applications
we can neglect the nuclear binding energy and the difference
between the proton and neutron masses and set $M=A m_p$. For
diatomic molecules the effective atomic number is related to the
reduced mass
\[
A_R = \frac{A_1A_2}{A_1+A_2}\;.
\]
Being a number, it cannot vary with time and can be dropped out
from any scaling equations applied for interpretation of
search-for-variation experiments.

Estimation of the rotational energy is also simple. The energy is
of order $L^2/I$ where $L$ is the (dimensional) orbital momentum and $I$ is the
moment of inertia. In atomic units $L$ is of order of unity and
$I\sim M R_0^2 \sim M$. Finally we find that the rotational energy
scales as $(1/M)$ in the atomic units or $(m_e/M)cR_\infty$ in the SI
units.

The importance of different scaling of the molecular transitions
was pointed out in \cite{thompson-99} due to astrophysical
applications. We summarize the scaling behavior of various
molecular transitions in Table~\ref{t:scal-m}.

\begin{table}
\begin{center}
\begin{tabular}{l|r}
\hline
Transition & Energy scaling \\
\hline
Electronic structure & $ cR_\infty $ \\
Vibrational structure & $(m_e/M)^{1/2} cR_\infty $ \\
Rotational structure & $(m_e/M) cR_\infty$ \\
\hline
\end{tabular}
\end{center}
\caption{Scaling behavior of the molecular energy intervals as
functions of the fundamental constants.  $M$ stands for an
effective nuclear mass (equal to reduce nuclear mass for diatomic
molecules), but for most of applications may be substituted for
the proton mass $m_p$. \label{t:scal-m}}
\end{table}

This evaluation shows a great convenience of the atomic units
for atomic and molecular calculations and demonstrates that a
calculation of the order of magnitude of an effect and its rough
detail is an important issue and a part of `calculable' properties.

At the present time all these scalings are not very hopeful for
laboratory searches since only two kinds of atomic transitions,
optical and HFS, are studied with a high accuracy. It should have a very
reduced application,
if the non-relativistic scaling in Tables~\ref{t:scal-a}
and~\ref{t:scal-m} would be the only method we have.

A successful deduction of constraints on a possible time variation
of $\{cR_\infty\}$ and $\alpha$ is possible because of the
relativistic corrections, which are responsible for a different
sensitivity to the $\alpha$-variation for various transitions of
the same kind (e.g., for various gross-structure optical
transitions). That was first pointed out in \cite{prestage-99} and
later successfully developed and applied to various atomic systems
in \cite{dzuba1-99}.

Most of standards deal with neutral atoms and single-charged ions.
The valent electron(s) spends most of time outside of a core
created by the nucleus and the closed shells. The core charge for
atoms used in the actual clocks is from one to three. The
relativistic correction is of order of $(Z^\prime\alpha)^2$. If
$Z^\prime$ is the core charge, it is still very small. However,
the relativistic corrections are singular and a contribution of
the short distances, where an electron sees the whole nuclear
charge $Z$, is enhanced. The dominant part of the relativistic
corrections comes from the short distances where the electron sees
the whole charge of a bare nucleus and thus in some atoms under
question the correction can be really big.

\subsection{Current laboratory limits}

Optical measurements delivered us data on a few elements (mercury
ion \cite{hg-99}, hydrogen \cite{h-99}, ytterbium ion
\cite{ybnew-99} and calcium \cite{ca-99}) and there are also
promising results on strontium ion \cite{strplus} and neutral
strontium \cite{stron} and more data on strontium and other
transitions are expected. The already available data
\cite{hg-99,h-99,ybnew-99,ca-99} are related to transitions with
very different relativistic corrections and that is enough to
derive strong limitations on the time variation of several
constants. The model-independent constraints achieved this way are
collected in Table~\ref{t:constr} (in the top part) \cite{AMOP}.
The HFS results were also applied to obtain constraints on the
variation of the magnetic moments \cite{rb-99,ybhfs-99}. To derive
results on the more fundamental quantities than the nuclear
magnetic moments of few particular nuclei, we applied the Schmidt
model (see, e.g., \cite{Schmidt}). Its importance for the
interpretation of results on the variation of the fundamental
constants was pointed out in \cite{Karshenboim-99}. The
model-dependent results are shown in the bottom part of
Table~\ref{t:constr} \cite{AMOP}.

\begin{table}
\begin{center}
\begin{tabular}{c|r}
\hline
Constants ($X$) & \multicolumn{1}{c}{Variation rate ($\partial \ln X / \partial t$)}  \\
\hline
$\alpha$ & $(-0.3\pm 2.0)\cdot 10^{-15}\,{\rm yr}^{-1}$\\
$\{cR_\infty\}$ & $(-2.1\pm 3.1)\cdot 10^{-15}\,{\rm yr}^{-1} $\\
$\mu_{\rm Cs}/\mu_{\rm B}$ & $(3.0\pm 6.8)\cdot 10^{-15}\,{\rm yr}^{-1} $\\
$\mu_{\rm Rb}/\mu_{\rm Cs}$ & $(-0.2\pm 1.2)\cdot 10^{-15}\,{\rm yr}^{-1} $\\
$\mu_{\rm Yb}/\mu_{\rm Cs}$ & $(3\pm 3)\cdot 10^{-14}\,{\rm yr}^{-1} $\\
\hline
$m_{\rm e}/m_{\rm p}$ & $(2.9\pm 6.2)\cdot 10^{-15}\,{\rm yr}^{-1} $\\
$\mu_{\rm p}/\mu_{\rm e}$ & $(2.9 \pm 5.8)\cdot 10^{-15}\,{\rm yr}^{-1} $\\
$g_{\rm p}$ & $(-0.1\pm 0.5)\cdot 10^{-15}\,{\rm yr}^{-1} $\\
$g_{\rm n}$ & $(3\pm 3)\cdot 10^{-14}\,{\rm yr}^{-1} $\\
\hline
\end{tabular}
\end{center}
\caption{Current laboratory constraints on the possible time
variations of natural constants \protect\cite{AMOP}. The results
above the horizonal line are model-independent, while the validity of the results
below the line depends on the applicability of the Schmidt model. The
uncertainty of this application is not shown.
 \label{t:constr}}
\end{table}

Is it possible to reach a model-independent constraint on the time
variation of of $m_e/m_p$ from atomic spectroscopy? Yes, it may be
done in the following way. First, we extract a limitation on a
variation of $cR_\infty$ without any use of the hydrogen data and
next we can compare it to a variation of the hydrogen $1s-2s$
frequency (which is proportional to $cR_\infty(1-m_e/m_p)$). The
constraint on the variation is
\begin{equation}
\frac{\partial \ln(m_p/m_e)}{\partial t} = (-0.4\pm 1.3)\cdot
10^{-11}\,{\rm yr}^{-1}\;,
\end{equation}
which is more than three orders of magnitude weaker than the
model-dependent constraint in Table~\ref{t:constr}. Stronger
model-independent limitations should appear rather from molecular
spectroscopy.

It is significant that we can eventually constrain the variability
of the fundamental constants. Results on variations of such
non-fundamental objects as the atomic transition frequencies
should be rather doubtful since such a level of accuracy has been
never achieved before and various details of the experiments may
need an additional examination. Expression such results in terms
of fundamental constants allows a cross-comparison and makes the
results more reliable.

\subsection{Non-laboratory searches for the variations of the
constants\label{ss:nolab}}

\begin{flushright}
\begin{minipage}{4cm}
... but it all came different!\\
\centerline{\em L.C.}
\end{minipage}
\end{flushright}

The laboratory limitations are not the strongest, but the most
reliable. Astrophysical \cite{astro} results unfortunately
contradict each other, as well as and geochemical \cite{oklo}
constraints. Studies by both methods involve various systematic sources
and the access to data is limited. One should deal with observations,
not with experiments.

\section{Fundamentality of the Constants and the Planck Scale\label{s:planck}}

\begin{flushright}
\begin{minipage}{9cm}
... and noticed that what can be seen from the old room was quite
common and uninteresting, but that all the rest was as different
as possible.\\ \centerline{\em L.C.}
\end{minipage}
\end{flushright}

If physics is governed by the most fundamental constants and if
the ultimate theory includes quantum properties of the space-time,
then the fundamental scale is determined by the Planck units
\begin{eqnarray}
M_{\rm Pl}&=& \left(\frac{\hbar c}{G}\right)^{1/2} =2.176\,45(16)\times10^{-8}\;\mbox{\rm kg}\nonumber\\
&~&~~~~~~~~~~~~~~~~~=1.220\,90(9)\times10^{19}\;\mbox{\rm Gev}/c^2\;,\nonumber\\
l_{\rm Pl}&=& \frac{\hbar}{M_{\rm Pl}\,c} =1.616\,24(12)\times10^{-35} \;\mbox{\rm m}\;,\nonumber\\
t_{\rm Pl}&=&\frac{l_{\rm Pl}}{c}=5.391\,21(40)\times10^{-44}\;\mbox{\rm s}\;,\nonumber\\
T_{\rm Pl}&=& \frac{M_{\rm Pl}c^2}{k}
=1.416\,79(11)\times10^{32}\;\mbox{\rm K}\;.
\end{eqnarray}
At the scale determined by these units the laws of Nature should
take the simplest form and if any observable fundamental
constant is calculable, it should be calculable there.

Due to success of the renormalization approach it is commonly
believed that physics from the Planck scale does not affect our
`low-energy' world. That is true only in part. To be accurate, the
idea of the renormalization reads that all what we need from the
higher-energy physics can be successfully measured at our low
energies. Still, what we measure at our energies comes from the
higher scale (such as, e.g., the Planck scale, or a scale of the
spontaneous violation of a larger symmetry related to the
unification). Presently, we do not have any theory related to the
higher energy scale. If the Planck or another high-energy scale
has no dynamics, we have not much hope to understand about the
high-energy physics from our low-energy experiments. We can learn
nothing from measured numbers until we are able to proceed with a
theory from the scale of, let us say, the $Z$ boson mass, to a
really high energy. Any corrections beyond that are small as
$(m/M_{\rm Pl})^2$ where $m$ is a characteristic mass scale we
deal with. However, if the Planck-scale physics has a dynamics,
e.g., a variation of certain parameters, that is not true anymore.
First of all, if the bare constants (such as the bare electron
charge $e_0$ and the the electron mass $m_0$) determined at the
Planck scale can vary, we should be able to detect that. There is
a chance, that the fine structure constant at the Planck scale is
calculable and, e.g., $\alpha_0 = 1/\pi^{4}$, but there is no
chance that any numerical exercises, such as done in the past, may
succeed to express actual $\alpha$ in an simple way. Neither it is
likely that $m_0$ is calculable in a simple matter. Nevertheless,
if the bare constants do not vary, we still can expect that the
dressed constant (i.e. the actual renormalized constants which we
measure) show a certain detectable variation, which could appear
via the renormalization.

How easily can we see such a dynamics, induced by the
renormalization? A question for $\alpha$ variation is whether the
divergencies are cut at the Planck scale, or a certain
supersymmetry enters into the game at an intermediate scale
$M_{\rm SS}\ll M_{\rm Pl}$ and cut the divergencies off. The other
question is whether the bare electron mass vary, and if it does
whether the ratios $m_e/M_{\rm Pl}$ and $M_{\rm SS}/M_{\rm Pl}$
vary.

In the case of ultraviolet divergencies going up to $M_{\rm Pl}$, the
result is
\begin{equation}
\frac{1}{\alpha}\,
\frac{\partial \alpha}{\partial t}
\sim \frac{\alpha}{\pi} \,\frac{1}{M_{\rm Pl}}\,
\frac{\partial M_{\rm Pl}}{\partial t}\;,
\end{equation}
while in the case of the supersymmetrical cut off, the variations
may be of a quite reduced value:
\begin{equation}
\frac{1}{\alpha}\,\frac{\partial \alpha}{\partial t} \sim \frac{\alpha}{\pi}\,
\left(\frac{M_{\rm SS}}{M_{\rm Pl}}\right)^2
\,\frac{1}{M_{\rm Pl}}\,
\frac{\partial M_{\rm Pl}}{\partial t}\;.
\end{equation}

Concerning the bare electron mass, a problem is much more
complicated, because detail of the so-called Higgs sector are
unclear and connection of the Higgs parameters with more
fundamental quantities are to be clarified. It may happen that a
dynamics at the Planck scale urges certain variations at the Higgs
sector and afterwards a variation of the Higgs vacuum average $v$
and consequently of the bare value of the electron mass $m_0$. We
note, that such a variation may in principle keep the ratio
$m_0/M_{\rm Pl}$ constant and reduce the value of the $\alpha$
variation via the renormalization which mentioned above. A
detection of a variation of the electron mass is even less clear.
A variation of $m_e$ is not the question from the experimental
point of view, the question is a variation of $m_e/m_p$. Since the
origin of the electron and the proton masses are very different it
is hard to understand what can happen with their ratio.

Still with a number of problems to be solved the fundamental
constants give us a chance to study physics of a higher-energy
scale not available anyhow else.

\section{Constants of Cosmology\label{s:cosm}}

\begin{flushright}
\begin{minipage}{7.5cm}
It's a poor sort of memory that only works backwards.\\
\centerline{\em L.C.}
\end{minipage}
\end{flushright}

Constants of our universe as a whole give us another questionable
chance to study physics beyond our essential world. Such constants
as listed in Table~\ref{t:cosm} may offer us a unique opportunity
to learn about the early time of the universe and thus perhaps about a
very-high-energy physics.

\begin{table}[hbtp]
\begin{center}
\begin{tabular}{c|c}
  \hline
Constant & Value  \\
 \hline
$\Omega_{\rm tot}-1$ & 0.02(2) \\
$\Omega_{\rm bm}$  & 0.044(4)  \\
$\Omega_{\rm dm}$  & 0.22(4)  \\
$\Omega_{\Lambda}$  &  0.73(4) \\
$T_{\rm CMB}$ & 2.725(1) K \\
$H$ & $ 0.73(4)\times10^{-10}\;{\rm yr}^{-1}$\\
$n_\gamma/n_B$ & $1.64(5)\times10^9$ \\
 \hline
\end{tabular}
\end{center}
\caption{Fundamental constants of cosmology. The results are taken
from \cite{PDG}.\label{t:cosm}}
\end{table}

In particular, these constants characterize the density of the bright matter (i.e.
a visible matter), the dark matter (a matter recognized because of its
gravitational effects) and the dark energy (recognized due to its
cosmological consequences and related to the Einstein's
$\Lambda$-term) in the units of the critical density
\begin{eqnarray}
\Omega_i &=& \frac{\rho_i}{\rho_c}\;,\\
\rho_c &=& \frac{3}{8\pi}\frac{H^2}{G}\;,
\end{eqnarray}
where $H$ is the Hubble constant. The critical density is a
separation mark between the closed ($\Omega_{\rm tot}>1$) and open
($\Omega_{\rm tot}<1$) universe. The between case ($\Omega_{\rm
tot}=1$) is the flat universe. The present result for the total
density is close to the flat value.

One more parameter, a ratio of the number of the microwave background
photons and the baryons ($n_\gamma/n_B$) is important to learn about a
moment when the light split from the baryon matter.

Meanwhile, we have to remember about the evolution in
understanding of such important `constants' as the free fall
acceleration $g$ and the water density. The constants of our world
are not necessarily the truly fundamental constants since their
values might be taken by chance with the spontaneous breakdown of
certain symmetries.

\section{Physics at the Edge\label{s:edge}}

\begin{flushright}
\begin{minipage}{9cm}
... As she couldn't answer either question, it didn't much matter
which way she put it.\\ \centerline{\em L.C.}
\end{minipage}
\end{flushright}

A question of the constancy and the fundamentality of the
fundamental constants is certainly a question of new physics.
Studying the problem experimentally via, e.g., searching for the
variability of the natural constants we address this new physics.
Maybe that is not the best way to do that, however, we are
extremely limited now in what we can do. Never from the Newton's
time, we have been so badly suited for going forward. The physics
is in a deep crisis, despite that it looks like as a success. We
are able to explain nearly everything we {\em can\/} deal with.
Yes, we have some problems, but that is either because some
objects are too complicated, or because the involved interaction
is strong and we are not able to go from Hamiltonian to the
observable quantities. But that is normal. In a sense, that is not
a problem of fundamental physics, but of technology to apply it,
which indeed is also of great importance.

We have access to only a few problems related in different ways to the
fundamentally new physics:
\begin{itemize}
\item on details of the Higgs sector;
\item on the extension of symmetry from the Standard model to a
certain unification theory which, probably, involves a supersymmetry;
\item on the dark matter;
\item on the dark energy;
\item on quantum gravity and physics at the Planck scale.
\end{itemize}
To address them, we suffer from extreme shortage of information
and do not see a feasible way to reach more data soon. To
illustrate the problem we summarize in Table~\ref{t:fund} data,
important for new physics.

\begin{table}[hbtp]
\begin{center}
\begin{tabular}{c|c}
  \hline
Constant  & Comment \\
 \hline
 $m_H$ & should be within certain margins if the Higgs particle is elementary\\
CKM matrix &  unitarity would confirm the minimal Standard Model \\
lepton analog of CKM &  would constrain physics beyond the Standard Model; \\
& the structure of both matrices could give a hint to the flavor symmetry\\
$|q_e+q_p|$ & should be zero in the case of unification theories\\
$\tau_p$ & would constrain unification theories$^\star$ \\
$d_e$ & would constrain unification theories$^\star$ \\
$d_n$  &  would constrain unification theories$^\star$ \\
$\sin\Theta_W$ & would constrain unification theories$^\dag$ \\
 $m_\nu$&  would constrain physics beyond the Standard Model;  \\
&  the Majorana mass would confirm that $q_\nu=0$ \\
\hline
$\Omega_{\rm tot}-1$ & if the value is positive, the universe is closed,
\\
&
if negative, it is open, if zero, it is flat;\\
 &  a small value is expected due to the inflation model (IM); it will constrain IM\\
$\Omega_{\rm dm}$  &  important that the dark matter exists$^\ast$ \\
$\Omega_{\Lambda}$  &  important that the dark energy exists$^\ast$  \\
$\partial H/\partial t$ &  would constrain cosmological theory \\
 \hline
\end{tabular}
\end{center}
\caption{Fundamental constants of new physics. $^\star$ -- the
experimental level is already within theoretical margins and any
improvement of the limits is important; $^\dag$ -- unification
theory must predict its value at a certain high energy scale;
radiative corrections are needed to go down to low energy for
a comparison with experiment; $^\ast$ -- the exact value is not
important for a moment. \label{t:fund}}
\end{table}

\section{Dreaming about New Physics\label{s:dream}}

\begin{flushright}
\begin{minipage}{8.5cm}
And here I wish I could tell you half the things Alice used to
say, beginning with her favourite phrase `Let's pretend.'\\
\centerline{\em L.C.}
\end{minipage}
\end{flushright}

What should scientists do with the obvious lack of the data?
Different people do different things. Some develop a `real sector'
of physics, where certain problems are important, sometimes very
important, but not
`fundamentally' important. Some develop tools and technologies
which are needed to go further. Some search for new physics,
but the lack of information does not allow to understand where it
is better to look for. So this search is a kind of a search for a
treasure which does not necessarily exist.

Some dream. It is hard to qualify differently the
theoretical studies without any connection with experiment, i.e.,
with reality. Dreamers existed at all times. Sometimes
doers and thinkers put the dreamers into shadow, but they existed.

The shortage of the experimental data makes us to wonder whether
`purely' theoretical progress is possible. Our opinion is rather
negative. The creation of the Einstein's special relativity is
sometimes believed to be a perfect example of such a progress.
However, there is certain confusion in use of the word
`theoretical'. A `theory' may be a model, a hypothesis or a
framework of certain calculations completely supported by
experiment, i.e., a high-level kind of fitting.

The famous inconsistency of the Newton's mechanics and the
Maxwell's theory of electromagnetism was not a `purely
theoretical' problem. That was a conceptual disagreement between
two `theoretical fits' of a huge amount of experimental data. The
relativity principle in the former form was a part of the Newton's
mechanics. The Einstein's solution reproduced both theories:
Maxwell's (exactly) and Newton's (as an approximation at $v/c\ll
1$). It was also immediately confirmed by numerous experiments.
Later, Einstein tried to solve an inconsistency between two other
pieces of the theoretical description of then existed data --- his
fresh-backed relativity and the old-fashion Newton's gravity. In
contrast to special relativity, it was difficult to confirm
general relativity accurately and in detail. The progress in the
field had been quite slow for a long period until vitalized by
appearence of new data.

The first important steps of quantum mechanics were directly
inspired by various experimental data (for that time this theory
was too crazy to appear as a result of a `purely theoretical'
development) and its crucial statements were immediately checked
experimentally. When the immediate experiment was not possible,
the theoretical ideas were sometimes gloriously correct, but
sometimes completely wrong. An example of the wrong ideas was an
expectation that the proton should have the Dirac's value of the
$g$-factor, i.e., $g_p=2$ (see, e.g., \cite{rigden}). There was no
other way but experiment to check if the idea is wrong or correct
and they had to wait until the idea was confirmed.

Sometimes this kind of studies is just a waste of time, sometimes
an important step to future theories. We can never know. Such
important conceptions as antiparticles, the Majorana mass, the
Kaluza-Klein theory, and the Yang-Mills gauge field appeared as
purely theoretical constructions. The positron was discovered
shortly after its prediction by Dirac. The Majorana mass perhaps
will now find its application to neutrino. The Yang-Mills gauge
field theory is now a way to describe weak and strong interactions
and likely other interactions which should appear due to the
unification. Speculations on Kaluza-Klein theories have inspired a
lot of works on unifications of all interactions with gravity,
but, maybe, this problem will be solved in another way.

\section{Conclusions\label{s:sum}}

\begin{flushright}
\begin{minipage}{9cm}
`Are you animal -- or vegetable -- or mineral?' he said, yawning
at every word.\\ \centerline{\em L.C.}
\end{minipage}
\end{flushright}

In this short paper we have tried to present an overview of a
problem of the fundamental constants and various related
questions, including practical metrology and realizations of
the units, simple atoms and macroscopic quantum phenomena, variations
of the constants and physics at the Planck scale.

In the beginning of the paper we introduced the fundamental constants
as certain universal parameters of the most basic equations. We
have noted afterwards that we hardly understand their origin.
These parameters play nearly a mystic role. Such equations as
Maxwell's or Dirac's appeared as then top-fundamental summary of
our understanding of Nature. Clearly, such equations interpret the
behavior of certain objects in terms of the input parameters $c$, $h$,
$e$, $m_e$, $G$ etc, which as any input parameters should come
from outside the equations. The equations happen to be a benchmark
between understood (the shape of the equations) and non-understood
(the fundamental constants inside the equations). In a sense the
truly fundamental constants are the least understood part of the
best understood physics.

We still do not know where the fine structure constant $\alpha$
comes from and what should its value be; we still wonder what
the reason that the gravity is so much weaker than the
electromagnetic interaction is. The trace of the origin of the
fundamental constants is lost somewhere in Wonderland, which we
can, perhaps, see through the looking glass and try to guess about
the unseen part of the room...

\section*{Acknowledgements}

This work was supported in part by the RFBR under grants
03-02-04029 and 03-02-16843, and by DFG under grant GZ 436 RUS
113/769/0-1. The author gratefully acknowledges stimulating
discussions with L. B. Okun, Z. Berezhiani, S. I. Eidelman, V.
Flambaum, L. Hollberg, O. Kancheli, D. Kleppner, C. L\"ammerzahl,
E. Peik, D. Pritchard, and V. A. Shelyuto.


\end{document}